\documentclass[preprint,authoryear,10pt]{elsarticle}

\usepackage{verbatim}

 \usepackage{graphics}
\usepackage{amssymb}
\usepackage{pst-tree}%permet d'écrire en couleur DANS LE PDF OU LE PS pas dans le DVI !!!!!
\usepackage{ulem}

\usepackage{url}

\journal{Icarus}

\DeclareTextSymbol{\degre}{T1}{6}
\DeclareTextSymbol{\degre}{OT1}{23}

\begin{document}

\begin{frontmatter}



\title{Instability zones for satellites of asteroids. The example of the (87) Sylvia system}

 \author[label1]{Julien Frouard}
 \ead{(To be added) rc.unesp.br}
 \author[label2]{Audrey Comp\`{e}re}
 \address[label1]{Instituto de Geoci\^{e}ncias e Ci\^{e}ncias Exatas, UNESP - Univ. Estadual Paulista, Departamento de Estat\'{\i}stica, Matem\'{a}tica Aplicada e Computa\c{c}\~{a}o, Av.24-A, CEP 13506-900, Rio Claro, SP, Brazil} 

 \address[label2]{Namur Centre for Complex Systems, naXys, University of Namur, Rempart de la Vierge 8, 5000 Namur, Belgium}

%\author{Frouard J. \& Comp\`{e}re A.}
%\address{}

\begin{abstract}
We study the stability of the (87) Sylvia system and of the neighborhood of its two satellites. We use numerical integrations considering the non-sphericity of Sylvia, as well as the mutual perturbation of the satellites and the solar perturbation. Two numerical models have been used, which describe respectively the short and long-term evolution of the system. We show that the actual system is in a deeply stable zone, but surrounded by both fast and secular chaotic regions due to resonances. We then investigate how tidal and BYORP effects modify the location of the system over time with respect to the instability zones. Finally, we briefly generalize this study to other known triple systems and to satellites of asteroids in general, and discuss about their distance from mean-motion and evection resonances.

\end{abstract}

\begin{keyword}
Celestial mechanics \sep Satellites of asteroids \sep Resonances, orbital \sep Satellites, dynamics


\end{keyword}

\end{frontmatter}

% \linenumbers


\section{Introduction}

A large number of satellites of asteroids have been discovered since the discovery of the satellite Dactyl, thanks to the Galileo flyby of (243) Ida (\citealt{belton1996}). 
%As of today, there is 110 known systems (binary and triple), following the online database (\url{http://www.asu.cas.cz/~asteroid/binastdata.htm}) described by \cite{Pravec2007} and \cite{Pravec2011}
As of today, there is 206 known systems (binary, triple and quintuple), following the Johnston's archive online database\footnote{\url{http://www.johnstonsarchive.net/astro/asteroidmoons.html}} (see also the online database\footnote{\url{http://www.asu.cas.cz/~asteroid/binastdata.htm}} described by \cite{Pravec2007} and \cite{Pravec2011}) and it is believed that small binaries could represent a fraction of 15\% of the NEA population (\citealt{Margot2002}; \citealt{Pravec2006}).  Triple systems are rare and only nine known systems have been reported up to now in the entire Solar System.   

The dynamical evolution and formation mechanisms of these systems are highly dependent on the size ratio between the secondaries and the primary. If this ratio is very small, as in the case of the Main-Belt asteroid (243) Ida, the systems are similar to the classical dynamical problem of a massless satellite orbiting a planet (see for example \citealt{Kozai1959,Kozai1962}), this one being replaced by a possibly highly elongated ellipsoid (\citealt{Chauvineau1993}; \citealt{Scheeres1994}; \citealt{Scheeres1996}; \citealt{Compere2012a}). On the other hand, systems with similar size components, as the Near-Earth asteroid (66391) 1999 KW4, have to be described taking into account both their shapes and their rotations. A lot of studies have been realized on the expression of the full two-body problem and the study of its characteristics (\citealt{Maciejewski1995}; \citealt{Scheeres2002}; \citealt{Fahnestock2008}; \citealt{Boue2009}). Similarly, emphasis has been given during the past decade on the description of dissipative effects on binary systems, like tidal effects (\citealt{Mathis2009}; \citealt{Goldreich2009}; \citealt{Taylor2010,Taylor2011}) or BYORP (\citealt{Cuk2005}; \citealt{Cuk2010}; \citealt{Mcmahon2010}; \citealt{Steinberg2011}).

%Jacobson \& Scheeres (2011) tides + byopr

We studied in this paper the dynamics and stability of the system (87) Sylvia, which was the first triple asteroid system discovered (\citealt{Marchis2005a}). The specificities of this system place it in the first class described above. Sylvia, discovered in 1866, is a low-eccentric and midly-inclined asteroid located in the outer Main Belt. Its long-term evolution has been investigated through the AstDys project (\citealt{Milani1998}; \citealt{Knezevic2003}) giving its proper orbital elements ($\overline{a}$ = 3.486 AU, $\overline{e}$ = 0.0537, $\overline{i}$ = $9.85^{o}$) and its secular fondamental frequencies ($n=55.297^{o}$/yr, $g = 134.798$"/yr, $s = -130.782$"/yr). Its orbit has been found to be slightly chaotic, exhibiting a Lyapunov time of $\sim$ 1.4 Myr.

The two satellites of Sylvia present near-circular and near-equatorial orbits, and have a mass ratio of about $10^{-4}$ and $10^{-5}$ with Sylvia. The outermost satellite, Romulus, is approximately ten times more massive than the innermost one, Remus, for a semi-major axis twice as important.
%and its semi-major axis is twice the one of Remus. 
\cite{Winter2009} studied the system and found that the satellites could be highly unstable when the oblateness of Sylvia (even a small fraction) is not taken into account. Indeed, the oblateness of the asteroid, as well as the short distance of the satellites from its surface ($\sim$ 5 and 10 radius of Sylvia), critically increase the precession frequencies and prevent them from commensurabilities with frequencies arising from other gravitational perturbations. 

Our aim is the understanding of the dynamical mechanisms present in the system and in its neighborhood. We then generalize some of the results to the other triple systems, and, in a general way, to the systems similar to (87) Sylvia, e.g. with a small size ratio and a primary diameter of the order of $\sim$ 100 km. 

\section{Study of the (87) Sylvia system}
 
The gravitational potential of Sylvia is modeled by a spherical harmonics expansion (e.g. \citealp{Kaula1966}):
\begin{equation}
U(r,\lambda,\phi)=-\frac{\mu}{r}\,\sum_{n=0}^\infty\, \sum_{m=0}^n \;\left(\frac{R_e}{r}\right)^n \; P_{n,m}(\sin \phi) \;\Big[C_{n,m} \, \cos (m \lambda) \, + \, S_{n,m} \, \sin (m \lambda)\Big],
\end{equation}
where $\mu$ is the gravitational constant of the central body, $R_e$ is its radius, ($r,\lambda,\phi$) are the spherical coordinates of the satellite, $C_{n,m}$ and $S_{n,m}$ are physical constants depending on the shape of the main body and named coefficients of the expansion ($n$ is the degree and $m$ is the order of the coefficients), and $P_{n,m}$ are the associated Legendre polynomials. 

The coefficients of this expansion are computed using the freely available software archive SHTOOLS\footnote{Available at \url{http://www.ipgp.fr/~wieczor/SHTOOLS/SHTOOLS.html}} developed by Mark Wieczorek and using the convex shape model of Sylvia (\citealt{Kaasalainen2002}; \citealt{Marchis2006}) available on DAMIT\footnote{This database is available at \url{http://astro.troja.mff.cuni.cz/projects/asteroids3D/web.php} (see \citealt{DAMIT} for more details)} (Database of Asteroid Models from Inversion Techniques). The asteroid shape models are represented by polyhedrons with triangular surface facets. %The model of Sylvia %, shown in Fig.\ref{shapeSylvia}, 
%is based on the observations by \cite{Kaasalainen2002} and \cite{Marchis2006}. 
Using this shape model as an input of a home-made code (which call the functions SHExpandLSQ, MakeGridDH and CilmPlus of SHTOOLS), we computed the spherical harmonics coefficients of Sylvia up to the tenth degree and order. After having run test integrations, we conclude that a $4^{th}$ order and degree harmonics development is sufficient to precisely approximate the perturbation due to the shape of Sylvia. The coefficients up to that degree and order are presented in Table \ref{coefHarmo}. More details on how they have been computed can be found in \cite{Compere2012b}.

In Table \ref{tabelem}, the orbital elements and some parameters used for the integrations are presented. The incertitudes (when known) are also given. The orbital elements and diameters of the satellites have been taken from \cite{Marchis2005a} and their masses from \cite{Winter2009}. The orbital elements of the two satellites from \cite{Marchis2005a} correspond to different epochs of reference, so we took as a reference epoch the mid-time between these epochs (JD 2453248) and move the mean anomalies of the satellites to this date, by considering fixed mean motions. The heliocentric orbital elements of Sylvia at this epoch, as well as its radius and rotation period, have been taken from the JPL service. The mass of Sylvia was obtained from \cite{Marchis2005a} and the ecliptic coordinates of its pole from \cite{Drummond2008}.
%\begin{figure}[t]
%\centering
%%\resizebox{12.cm}{!}{\includegraphics [angle=0,width=\textwidth] {shapeSylvia.eps}}
%\caption{Shape model of Sylvia available on DAMIT. This figure comes from the database DAMIT.}
%\label{shapeSylvia}
%\end{figure}
\begin{tiny}
\begin{table}
 \begin{center}
\begin{tabular}{ccrr}
Degree (n) & Order (m) & $C_{n,m}$ & $S_{n,m}$\\
\hline
    2  &  0 & $-0.1437660949676515$ & $0$\\
    2  &  1 & $ 0.3590487423556445 \times 10^{-3}$ & $-0.7369685217769826 \times 10^{-2}$\\
    2  &  2 & $-0.2103019269994283 \times 10^{-1}$ & $-0.4521411518503407 \times 10^{-1}$\\
    3  &  0 & $0.2652106240427706 \times 10^{-2}$ & $0$\\
    3  &  1 & $0.6643161063285490 \times 10^{-2}$ & $0.1521608001095167 \times 10^{-2}$\\
    3  &  2 & $-0.2228678182750910 \times 10^{-3}$ & $-0.3961687875836957 \times 10^{-3}$\\
    3  &  3 & $-0.1532096089753877 \times 10^{-2}$ & $0.3681625131424677 \times 10^{-2}$\\
    4  &  0 & $0.3305186797474446 \times 10^{-1}$ & $0$\\
    4  &  1 & $-0.3043347534177666 \times 10^{-2}$ & $0.5784390224634026 \times 10^{-2}$\\
    4  &  2 & $-0.4028484233892346 \times 10^{-3}$ & $0.3763942276631470 \times 10^{-2}$\\
    4  &  3 & $-0.2777959861061120 \times 10^{-4}$ & $-0.1436540399555624 \times 10^{-3}$\\
    4  &  4 & $-0.2636496691209342 \times 10^{-3}$ & $0.8129584824760719 \times 10^{-4}$\\
\end{tabular}
\end{center}
\caption{Main spherical harmonics coefficients of Sylvia.}% computed using the software SHTOOLS and the shape model available on DAMIT.}
\label{coefHarmo}
\end{table} 
\end{tiny}
%\begin{tiny}
%\begin{table}
% \begin{center}
%\begin{tabular}{lll}
%\textbf{Sylvia} & & \\
%radius R & 130.47 km ($\sigma$=6.65)& JPL\\
%shape &  & Carry\\
% &  & Descamp\\
%mass M & 1.478 $\times$ $10^{19}$ kg& ?\\
%rotation period & 5.184 h & JPL \\
%phase &&\\
%orbital elements (JD 2453248)  & & JPL Horizon\\
%ecliptic coordinates of the pole&$\alpha$=100°, $\delta$=62°& Drummond \& Christou 2008\\ 
%\hline
%\textbf{Remus}&&\\
%semi-major axis&706 $\pm$ 5 km& Marchis et al. 2005\\
%eccentricity&0.016 $\pm$ 0.011& Marchis et al. 2005\\
%inclination&$2^o$ $\pm$ $1^o$& Marchis et al. 2005\\
%mean anomaly&$96^o$.087& Marchis et al. 2005\\
%argument of pericenter&$314^o$& Marchis et al. 2005\\
%longitude of node&$97^o$& Marchis et al. 2005\\
%mass & 2.154 $\times$ $10^{14}$ kg& ?\\
%diameter & 7 $\pm$ 2 km&Marchis et al. 2005\\
%\hline
%\textbf{Romulus}&&\\
%semi-major axis&1356 $\pm$ 5 km& Marchis et al. 2005\\
%eccentricity&0.001 $\pm$ 0.001& Marchis et al. 2005\\
%inclination&$1.7^o$ $\pm$ $ 1^o$& Marchis et al. 2005\\
%mean anomaly&$324^o$.308& Marchis et al. 2005\\
%argument of pericenter&$273^o$& Marchis et al. 2005\\
%longitude of node&$101^o$& Marchis et al. 2005\\
%mass & 3.6625 $\times$ $10^{15}$ kg& ?
%%diameter & 18 $\pm$ 4 km&Marchis et al. 2005\\
%\end{tabular}
%\end{center}
%\caption{Orbital elements, physical parameters and corresponding incertitudes}
%\label{tabelem}
%\end{table} 
%\end{tiny}
\begin{tiny}
\begin{table}
\begin{center}
\begin{tabular}{lll}
\textbf{Sylvia} & & \\
radius R & 130.47 km ($\sigma$=6.65)\\
mass M & 1.478 $\times$ $10^{19}$ kg\\
rotation period & 5.184 h \\
ecliptic coordinates of the pole&$\alpha$=100\degre, $\delta$=62\degre\\ 
\hline
\textbf{Remus}&&\\
semi-major axis&706 $\pm$ 5 km\\
eccentricity&0.016 $\pm$ 0.011\\
inclination&2\degre $\pm$ 1\degre\\
mean anomaly&96.087\degre\\
argument of pericenter&314\degre\\
longitude of node&97\degre\\
mass & 2.154 $\times$ $10^{14}$ kg\\
diameter & 7 $\pm$ 2 km\\
\hline
\textbf{Romulus}&\\
semi-major axis&1356 $\pm$ 5 km\\
eccentricity&0.001 $\pm$ 0.001\\
inclination&1.7\degre $\pm$ 1\degre\\
mean anomaly&324.308\degre\\
argument of pericenter&273\degre\\
longitude of node&101\degre\\
mass & 3.6625 $\times$ $10^{15}$ kg\\
diameter & 18 $\pm$ 4 km
\end{tabular}
\end{center}
\caption{Orbital elements, physical parameters and corresponding incertitudes of the bodies.}
\label{tabelem}
\end{table} 
\end{tiny}

Based on this shape model, we used two dynamical models to investigate the dynamics of the system. One was used for short-term and precise numerical integrations, while the second one was used to study the secular evolution of the system.   

\subsection{Short-term numerical integrations}

Our first set of integrations was based on the complete equations of the orbital motion of the satellites Remus and Romulus. These simulations were computed with the software NIMASTEP, presented in \cite{NIMASTEP}, which is a home-made numerical software. It allows to integrate the osculating motion (using Cartesian coordinates) of an object considered as a point mass orbiting a homogeneous central body which rotates constantly around its principal moment of inertia. It has been successfully tested and used in many papers, as for example \cite{Delsate2010}, \cite{Lemaitre2009}, \cite{Valk2009}, \cite{Delsate2011}, \cite{Compere2012a}. To the aim of this work, the software has been improved in order to integrate the motion of two interacting satellites.
% \citet{Delsate2010,Lemaitre2009,Valk2009,Delsate2011,Compere2011}. 

\begin{figure}[h]
\centering
%\resizebox{13.cm}{!}{\includegraphics [angle=0,width=\textwidth] {forcesSylvia-print2.eps}}
\resizebox{13.cm}{!}{\includegraphics [angle=0,width=\textwidth] {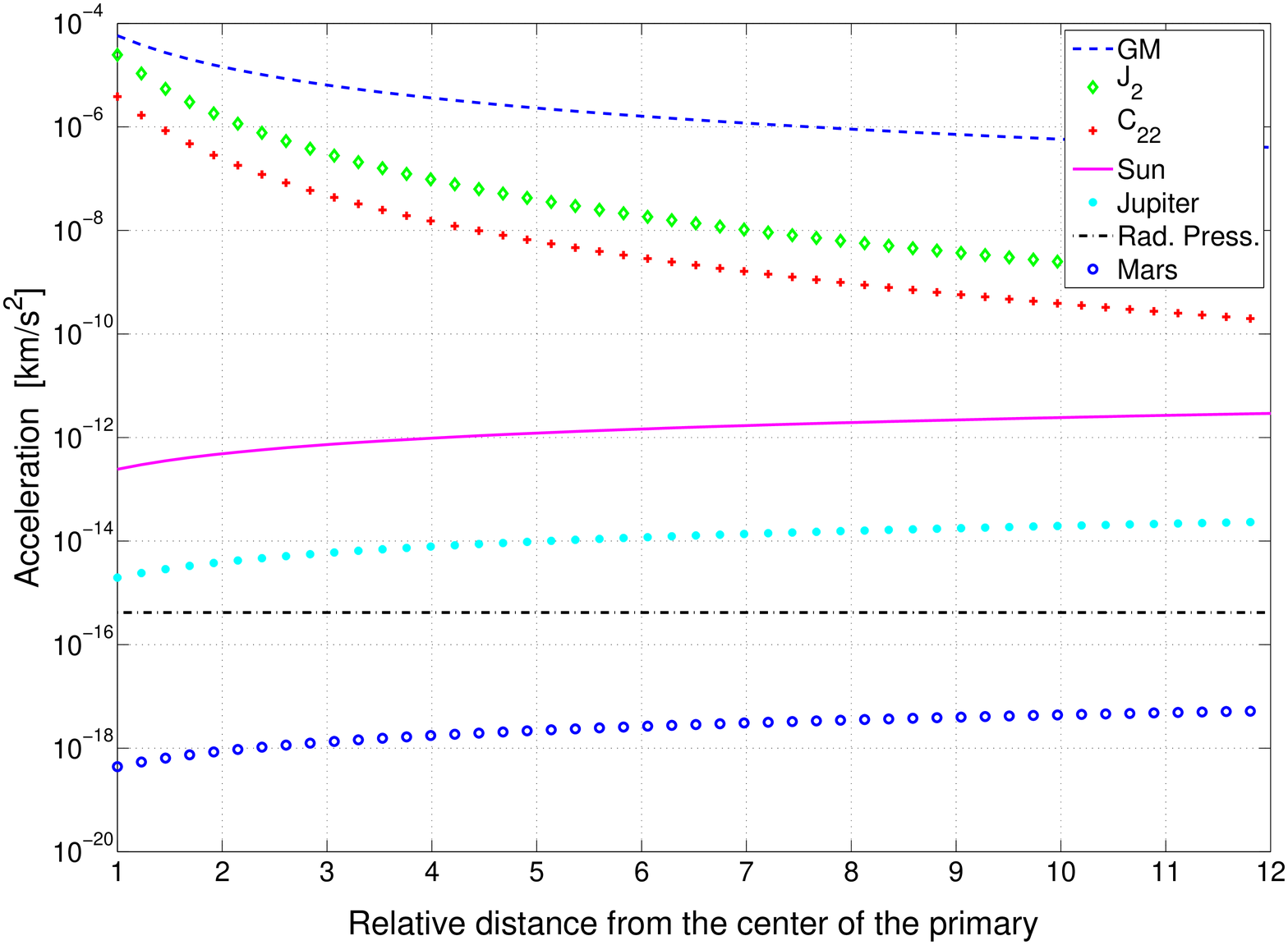}}
\caption{Order of magnitude of the accelerations of a satellite of Sylvia due to various perturbations, in function of the distance from Sylvia (expressed in radius of Sylvia).}
\label{fig10}
\end{figure}

The direct gravitational perturbations of the terrestrial and giant planets are negligible, as well as the orbital variations of Sylvia for the timespans considered here. We show in Fig.\ref{fig10} the importance of the two-body problem and of various perturbations on the acceleration of the satellite (only the radial component), in function of the relative distance from Sylvia, which is the ratio between the distance from the center of Sylvia and its equatorial radius. These results have been confirmed by numerical simulations. We consider in the following a Keplerian, eccentric and inclined orbit for Sylvia. The rotation of Sylvia is supposed to be constant and its axis of rotation, also constant, corresponds to its principal axis of inertia. 

We used the MEGNO indicator (\citealt{Cincotta2000}) in order to distinguish among regular and chaotic orbit. The MEGNO is a fast chaos indicator based on the numerical integration of a tangent vector, that has proved to be reliable in a large class of dynamical systems (\citealt{Cincotta2003}; \citealt{Gozdziewski2001,Gozdziewski2008a}; \citealt{Breiter2005}; \citealt{Compere2012a}; \citealt{Frouard2011}).  

We numerically integrated the expressions of \cite{Gozdziewski2001} along with the equations of motion. The initial tangent vector was chosen as random for each integrated orbit. Since the two satellites perturb each other, it is worth noting that the MEGNO indicator here was computed from the orbital evolution of the two satellites taken as a whole. The indicator thus represents the behavior of the complete system.

\begin{figure}[h]
\centering
%\resizebox{8.cm}{!}{\includegraphics [angle=0,width=\textwidth] {aRem-eRem.eps}}
%\resizebox{8.cm}{!}{\includegraphics [angle=0,width=\textwidth] {aRom-eRom.eps}}
\resizebox{8.cm}{!}{\includegraphics [angle=0,width=\textwidth] {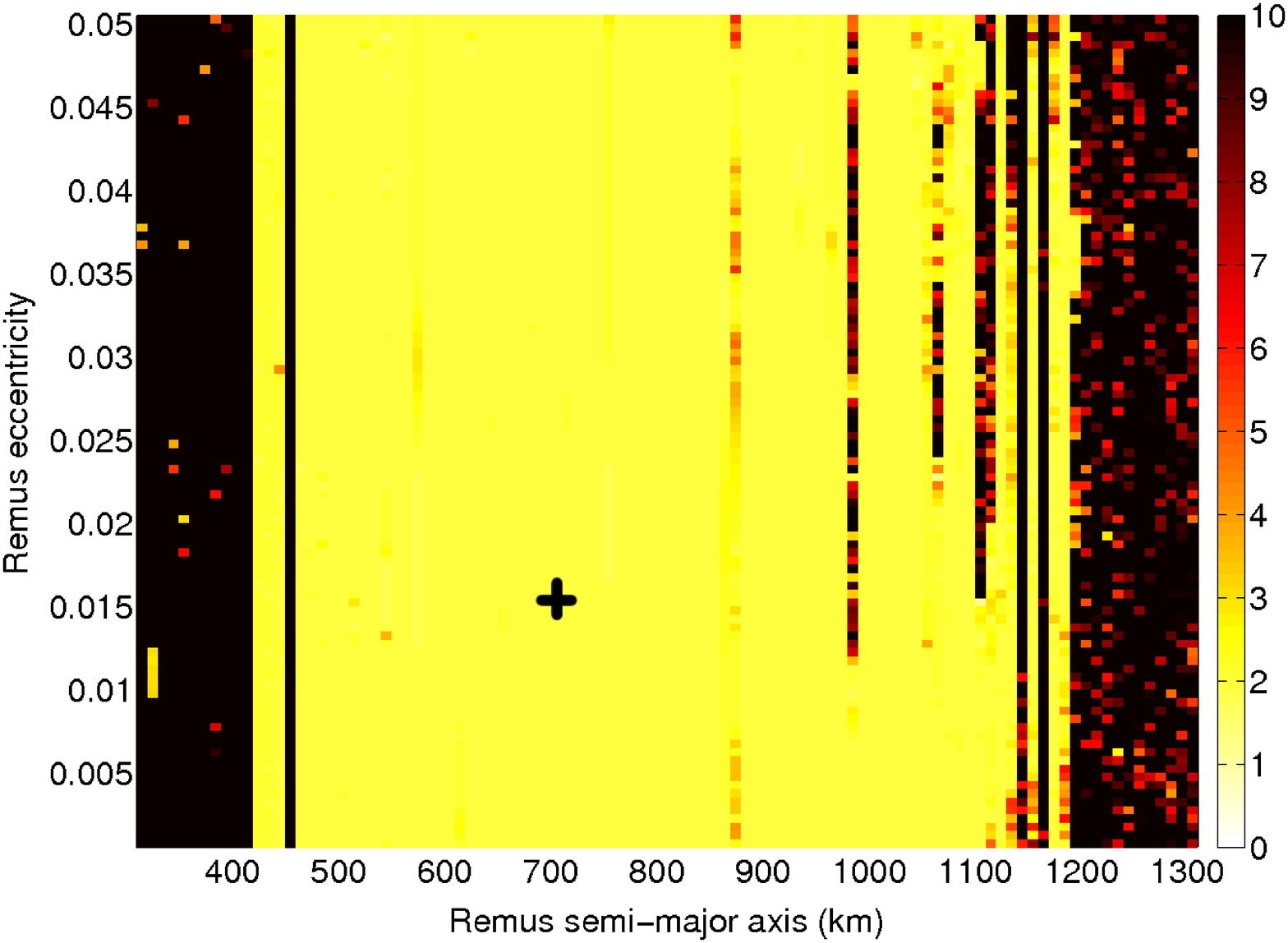}}
\resizebox{8.cm}{!}{\includegraphics [angle=0,width=\textwidth] {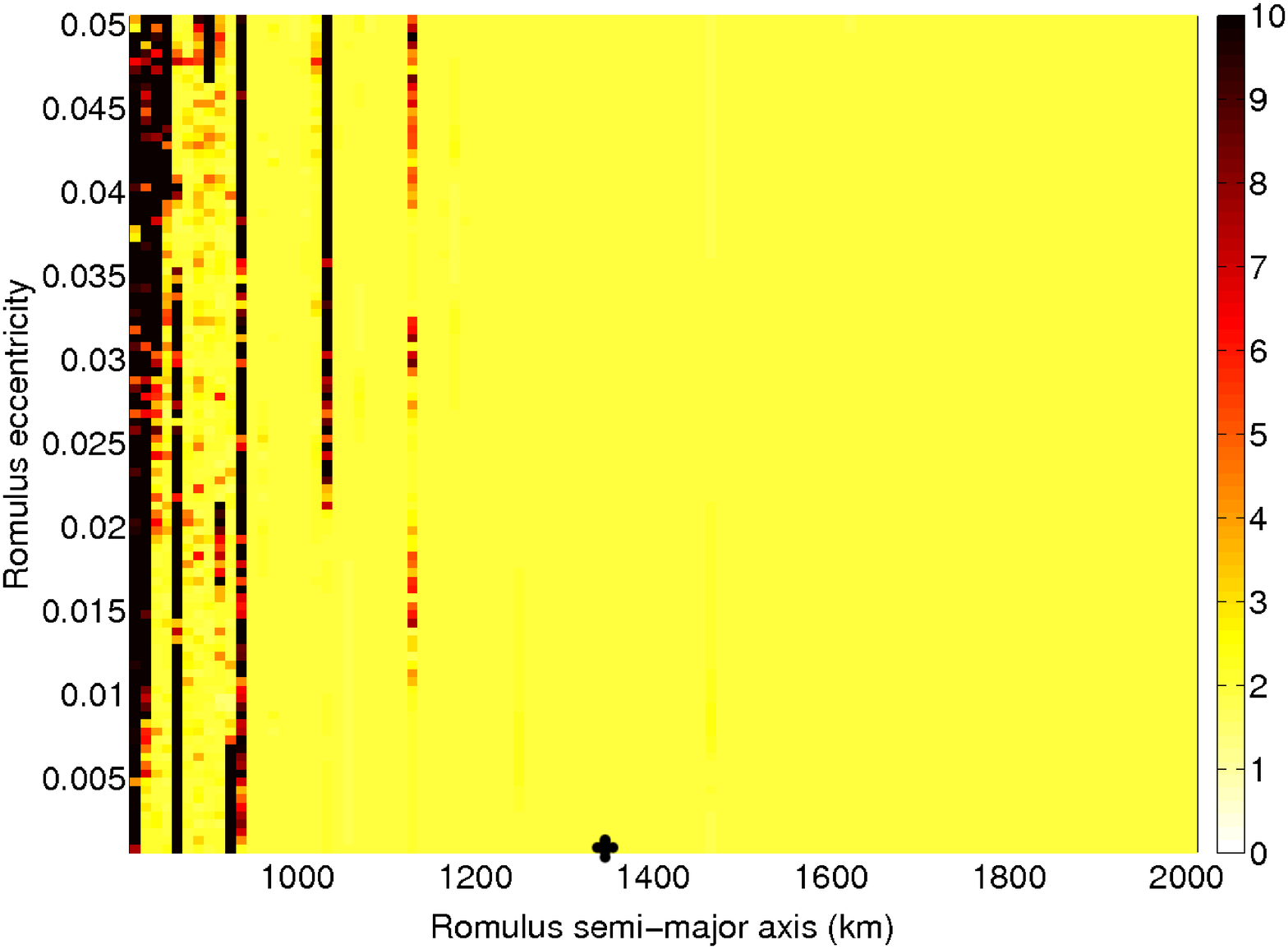}}
\caption{MEGNO map of Remus (top) and Romulus (bottom) in semi-major axis and eccentricity. Stable quasi-periodic orbits correspond to a color code of 2 (yellow), while chaotic orbits have a MEGNO $>$ 2 (orange to black). A black cross indicates the actual position of the Remus-Romulus system.}
\label{fig11}
\end{figure}

We integrated a large set of orbits by varying the semi-major axis and the eccentricity of the two satellites, while keeping all the other initial variables as constant. The numerical integrations had a timespan of 20 years, until the satellites collided each other or with Sylvia, or in the case of ejection into heliocentric orbits. The results of the integrations are shown in Fig.\ref{fig11}.  

The actual position of Remus and Romulus lies in a very stable zone, even when considering their known incertitudes in semi-major axis and eccentricity (see Table \ref{tabelem}). The map clearly shows two types of instability zones (in black in the figure). One appears systematically for semi-major axis of Remus under 400 km. Under this value, Remus quickly collides with Sylvia or is ejected from the system. 
%Although not shown in this figure, the same fate happens for Romulus, when its initial semi-major axis is taken under 400 km. 
%The satellites have initially very low values of the eccentricity, but satellites with initially more eccentric orbits are unstable for larger values of the semi-major axis, their orbit being destabilized by spin-orbit resonances. 
The satellites have very low initial values of the eccentricity, which locates the orbits far from regions where overlapping of spin-orbit resonances is important. The width and shape of spin-orbit resonances in the general case of an orbiter have been investigated by \cite{Mysen2006} and \cite{Mysen2007}. A narrow vertical line of instability can thus be seen for Remus, at $a_{rem}$ = 440 km, which corresponds to the spin-orbit resonance 3:1 (corresponding to the argument $3 \lambda_{rem} - \lambda_{spin}$, where $\lambda_{rem}$ is the longitude of Remus and $\lambda_{spin}$ the spin angle of Sylvia). Higher-order resonances, like the 4:1 at $a_{rem}$ = 530 km are too weak to destabilize the orbit. 

\begin{figure}[h]
\centering
%\resizebox{12.cm}{!}{\includegraphics [angle=0,width=\textwidth] {CarteRomRemus-aRomaRemus-10ans-megnoRemus.eps}}
%\resizebox{12.cm}{!}{\includegraphics [angle=0,width=\textwidth] {carte-haut-article.eps}}
\resizebox{12.cm}{!}{\includegraphics [angle=0,width=\textwidth] {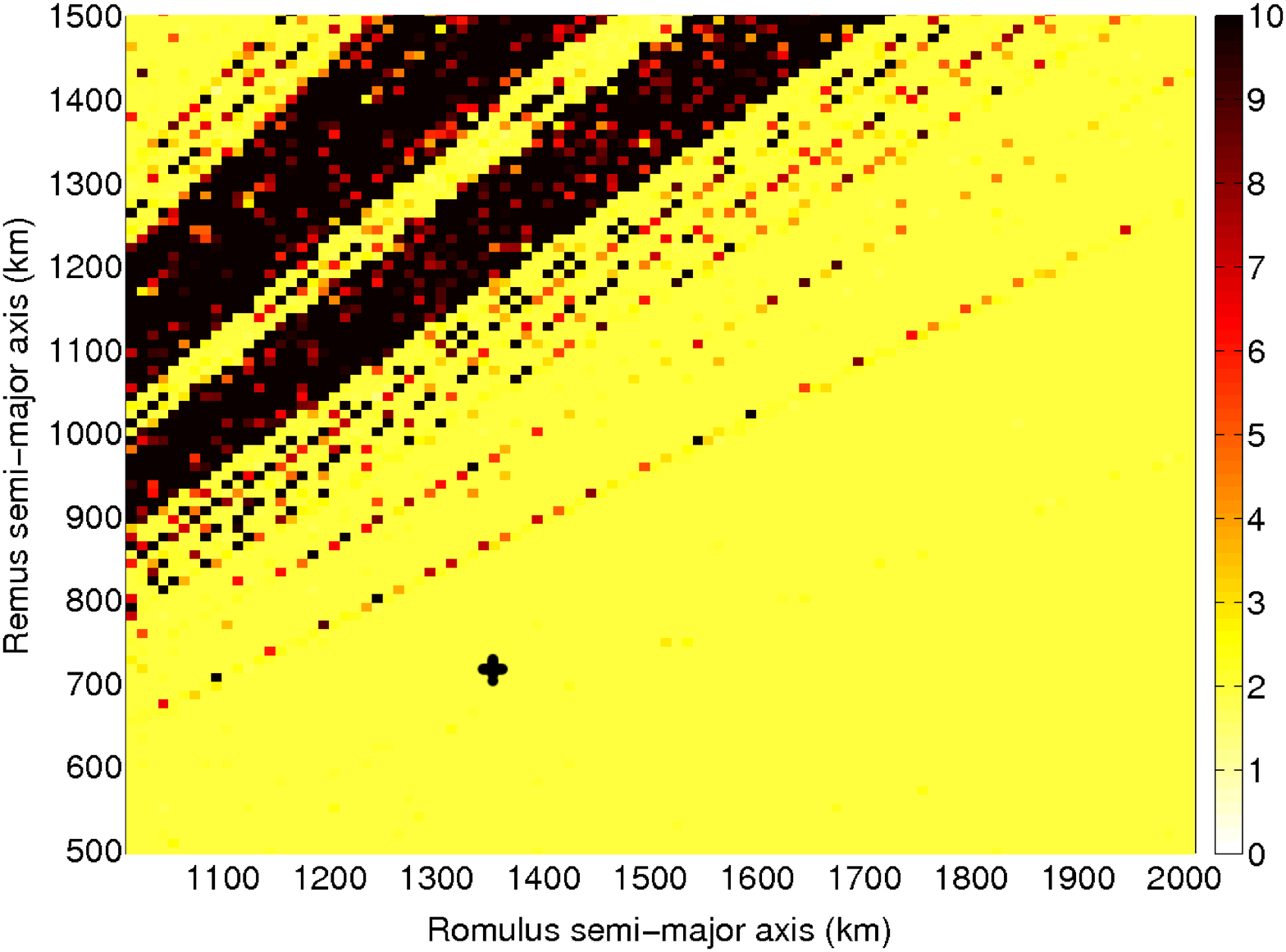}}
\caption{MEGNO map in semi-major axis of both satellites. A black cross indicates the actual position of the Remus-Romulus system.}
\label{fig1}
\end{figure}

The second type of instability zones corresponds to mean-motion resonances (MMRs) between the two satellites, at high semi-major axis in the Remus map, and low semi-major axis in the Romulus map. The MMRs have a typical V-shape (see for example \citealt{Gozdziewski2008b}; \citealt{Bazso2010}) which enlarges them for increasing eccentricity. These resonances can be seen clearlier in Fig.\ref{fig1}, where we modify the semi-major axis of the satellites in the range $500-1500$ km (Remus) and $1000-2000$ km (Romulus). A same approach has been applied to investigate the dynamics of the giant planets by \cite{Guzzo2005,Guzzo2006}. 
%We integrated $?\times?$ orbits for a timespan of 20 years, where the initial semi-major axis are taken 
%We integrated a large set of orbits by varrying the semi-major axis of the two satellites, while keeping all the other initial variables as constant. A same method has been applied to the giant planets by \cite{Guzzo2005,Guzzo2006}. We integrated $?\times?$ orbits for a timespan of 20 years, where the initial semi-major axis are taken in the range $1000-2000$ km (Romulus) and $200-1500$ km (Remus). The numerical integration is stopped when the satellites collide between themselves or with Sylvia, or when they are ejected in heliocentric orbits. The results of the integrations are shown in Fig.\ref{fig1}.  
%The map clearly shows two types of instability zones (in black in the figure). One happens systematically for semi-major axis of Remus under 400 km. Under this value, Remus quickly collides with Sylvia. Although not shown in this figure, the same fate happens for Romulus, when its initial semi-major axis is taken under ?? km. The satellites have initially very low values of the eccentricity but satellites with initially more eccentric orbit are unstable for larger values of the semi-major axis, their orbit being destabilized by spin-orbit resonances. The width and shape of spin-orbit resonances in the general case of an orbiter have been investigated by \cite{Mysen2006} and \cite{Mysen2007}.
%The satellites almost immediately collide with Sylvia, due to their initial low eccentricity. 
%The second instability zones correspond to mean-motion resonances (MMRs),

The MMRs correspond to the different lines, each one describing a resonance $k_1 n_1 \backsimeq k_2 n_2$, where $n_1$ and $n_2$ are the mean motions of the satellites, and $k_1$, $k_2$ integers. By neglecting the masses of the satellites with respect to the mass of Sylvia, the lines are described by the relation $a_2 \backsimeq (k_2/k_1)^{2/3} a_1$. 
%A strict resonance should satisfy the argument $k_1 n_1 + k_2 n_2 + k_3 \dot{\varpi}_1 + k_4 \dot{\varpi}_2 + k_5 \dot{\Omega}_1 + k_6 \dot{\Omega}_2 = 0$, thus creating multiplets for each mean motion resonances (Morbidelli 2002). However, as the eccentricites and inclinations of the satellites are very low, 
%The resonances accumulate and then overlap as the satellites are closer.
The resonances overlap as the satellites have closer semi-major axis. The two important instability zones present on each side of the $a_2 = a_1$ line represent initial orbits leading to close encounters between the satellites, resulting in collisions or ejections from the system after a chaotic evolution.

\begin{figure}[h]
\centering
\resizebox{12.cm}{!}{\includegraphics [angle=0,width=\textwidth] {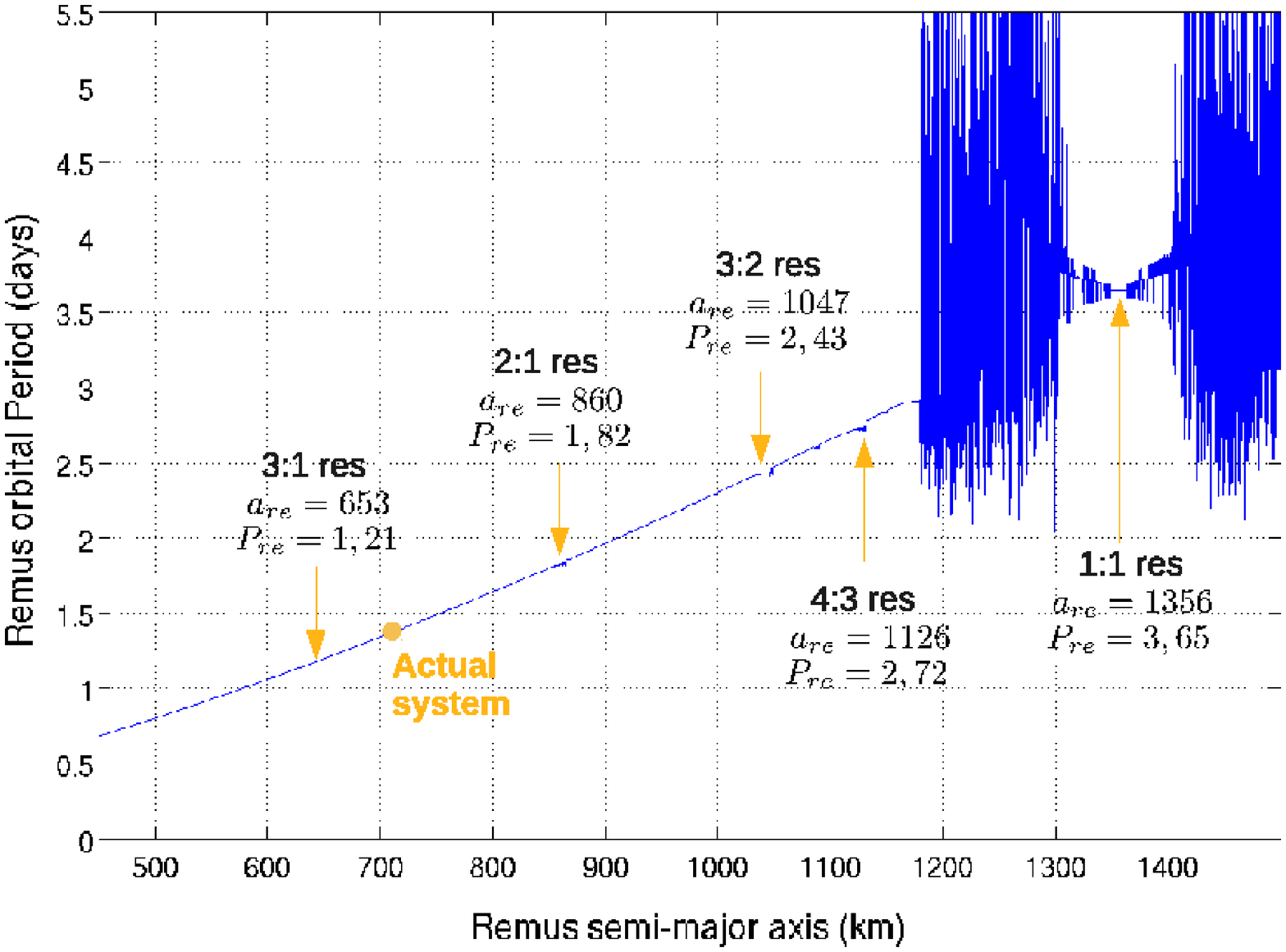}}
\caption{Period of Remus obtained from frequency analysis, and location of the main mean-motion resonances.}
\label{fig9}
\end{figure}

The actual position of the system is represented by a cross in the figure. The system thus lies in a very stable zone, bounded by the MMR 2:1 and 3:1, and shows a very regular evolution for at least $10^{4}$ years. %The Fig.\ref{fig3} shows a zoom of the Fig.\ref{fig1} around the actual position of the system. 
%Figure \ref{fig2} shows the maximum eccentricity attained by the satellites over 20 years, in function of the initial semi-major axis of Remus, while the initial semi-major axis of Romulus is equal to its actual value (thus investigating a vertical line in Fig.\ref{fig1}).
Figure \ref{fig9} shows the period of Remus in function of the initial semi-major axis of Remus, while the initial semi-major axis of Romulus is equal to its actual value (thus investigating a vertical line in Fig.\ref{fig1}). The period was obtained by a frequency analysis (\citealt{Laskar1990}; \citealt{Laskar2005}) of the complex time series ($a \cos \lambda,a \sin \lambda$).

\begin{figure}[h]
\centering
%\resizebox{6.cm}{!}{\includegraphics [angle=0,width=\textwidth] {eRemLong.eps}}
%\resizebox{6.cm}{!}{\includegraphics [angle=0,width=\textwidth] {eRomLong.eps}}
\resizebox{6.cm}{!}{\includegraphics [angle=0,width=\textwidth] {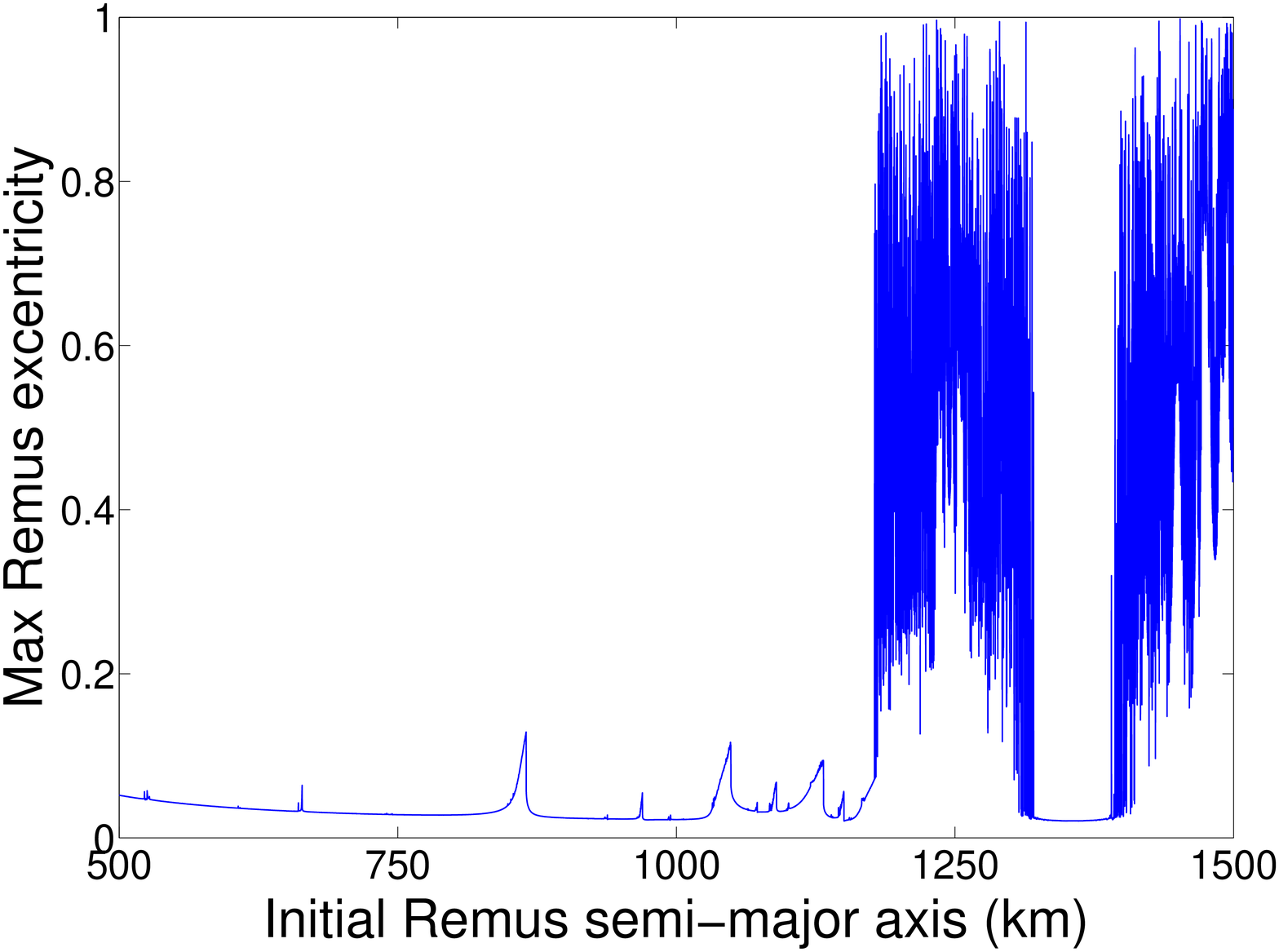}}
\resizebox{6.cm}{!}{\includegraphics [angle=0,width=\textwidth] {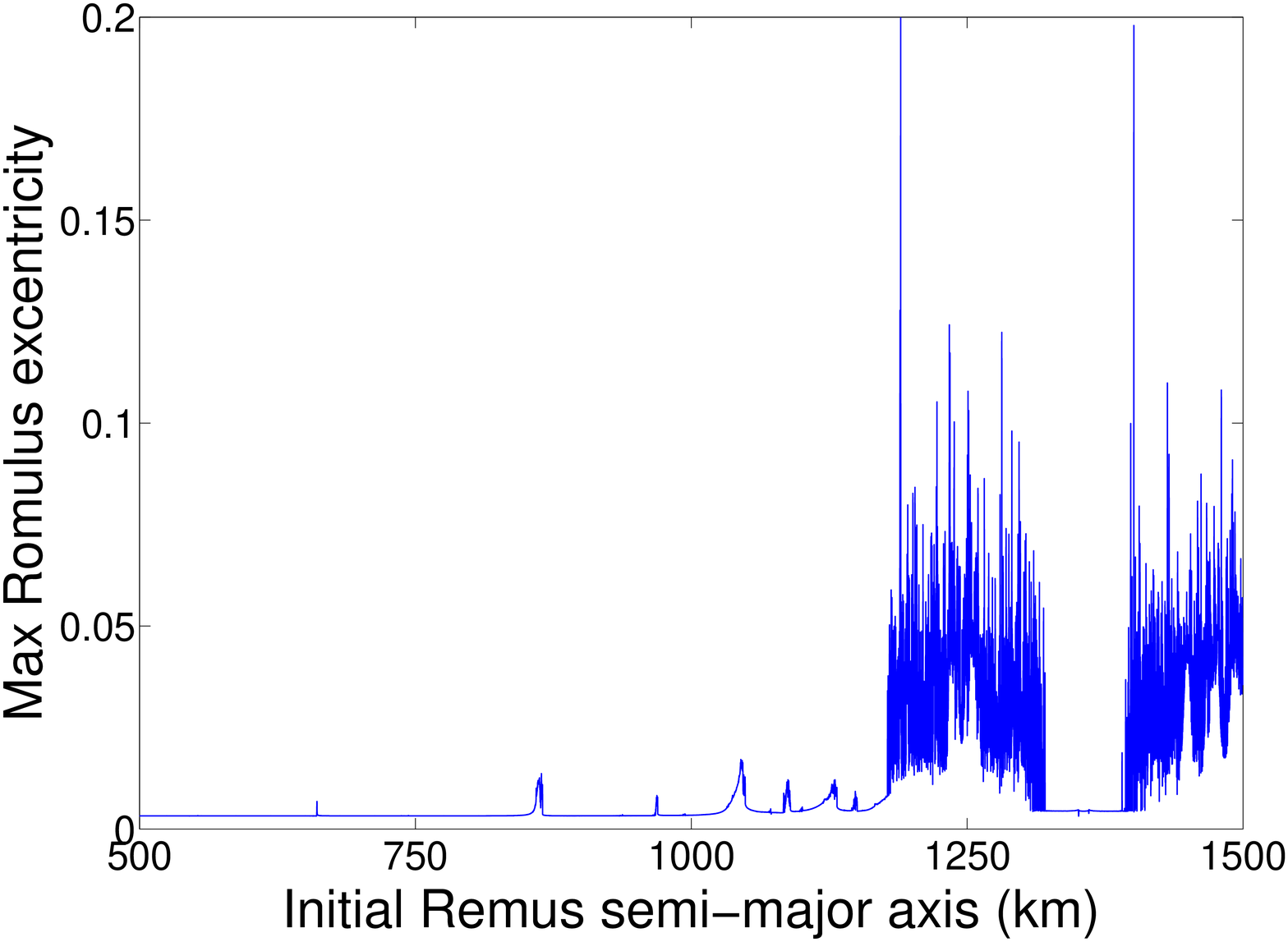}}
\resizebox{6.cm}{!}{\includegraphics [angle=0,width=\textwidth] {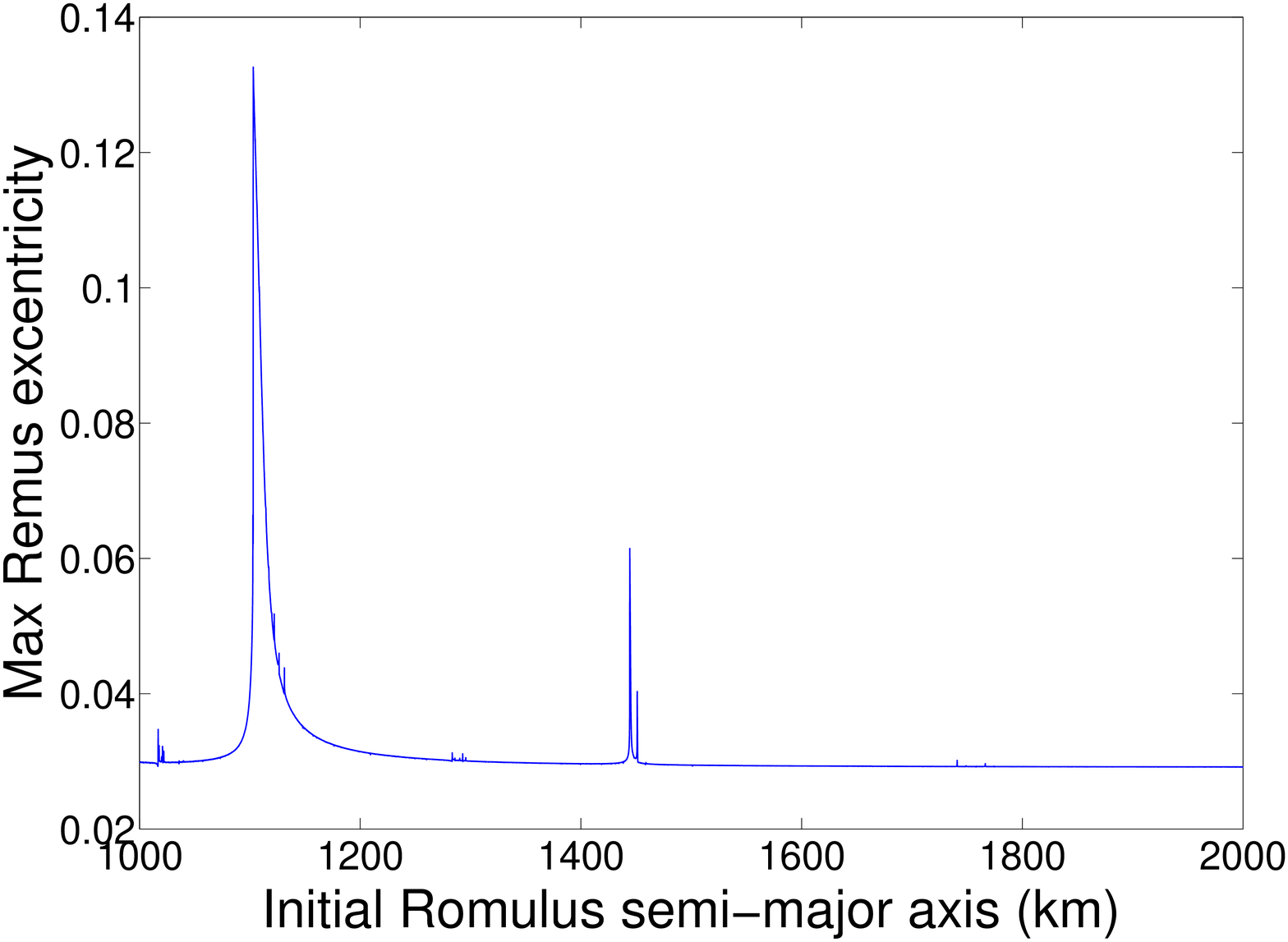}}
\resizebox{6.cm}{!}{\includegraphics [angle=0,width=\textwidth] {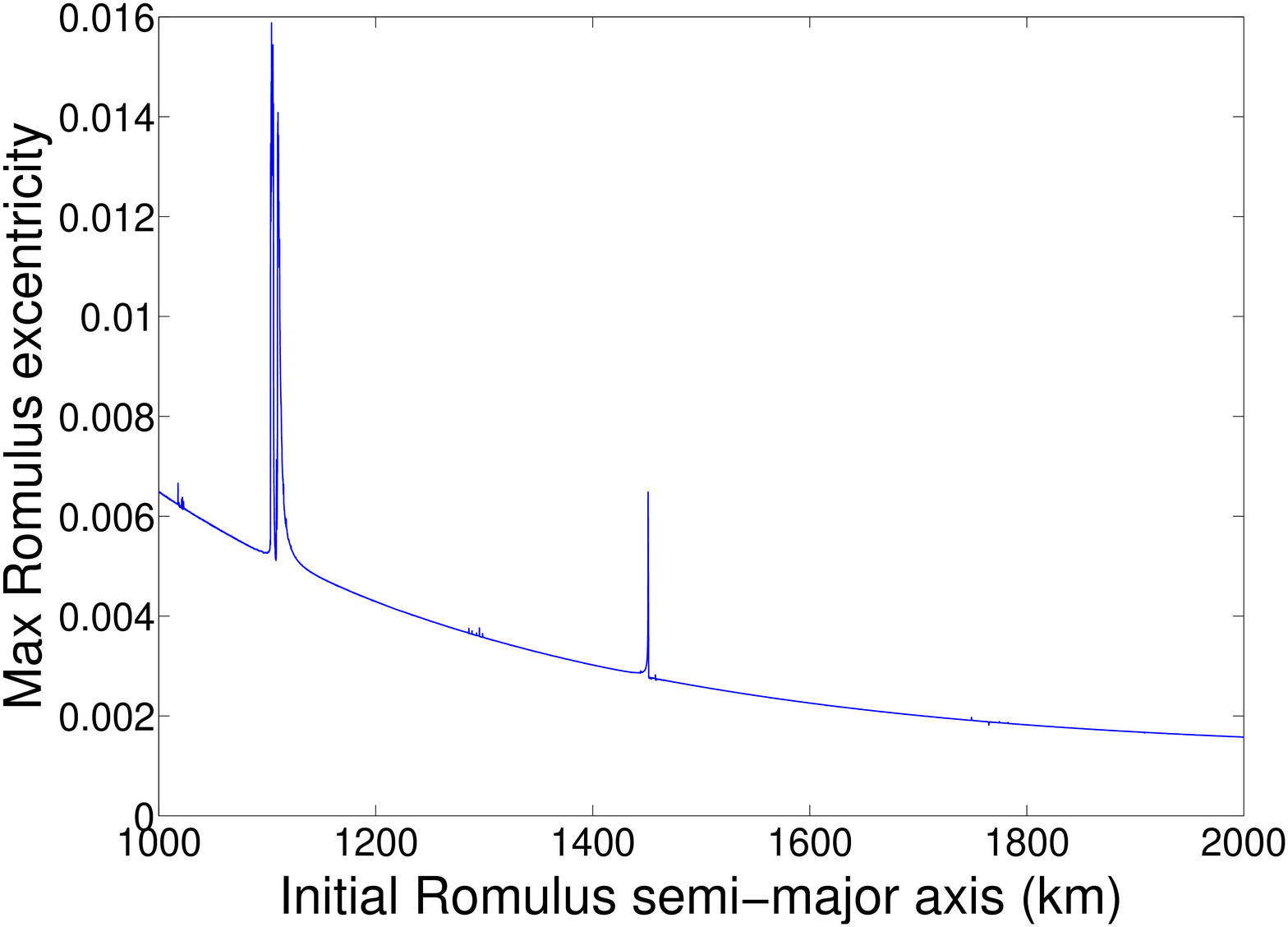}}
\caption{Maximum eccentricity attained by Remus (left) and Romulus (right) over 20 years, in function of the initial semi-major axis of Remus and Romulus. The two important peaks seen in the two bottom panels correspond to the MMR 2:1 and 3:1.}
\label{fig2}
\end{figure}

Figure \ref{fig2} shows the maximum eccentricity attained by the satellites over 20 years, in function of their initial semi-major axes. 
The figure indicates that the eccentricity of Remus can possibly raise up to relatively large values inside the major MMRs (up to 0.13 inside the MMR 2:1), and ten times more than the eccentricity of Romulus, which just reflects the fact that due to their respective masses, Remus is much more perturbed by Romulus than the reciprocal. 

We will now present the results obtained by investigating the secular evolution of the system.

\subsection{Long-term numerical integrations}

The main limiting factor in the numerical integrations of the above model concerning CPU time is the fast rotation period of Sylvia (5.184 hours). We thus used equations of motion averaged over the spin angle of Sylvia, as well as over the mean anomalies of the satellites, to study the behavior of the system over longer timescales.

We numerically integrated the Lagrange equations (\citealt{Brouwer1961}) where the averaged disturbing function $\langle R \rangle$ is the sum of different averaged perturbations $\langle R \rangle = \langle R_{mut} \rangle + \langle R_{obl} \rangle + \langle R_{\odot} \rangle$. The mutual perturbation between the satellites $\langle R_{mut} \rangle$ is approximated by a fourth-order expansion in eccentricity and inclination of their disturbing function (\citealt{Murray2000}). The oblateness perturbation $\langle R_{obl} \rangle$ is represented by an expansion of fourth-order in eccentricity and inclination of the oblateness disturbing function, containing terms in $J_2,J_2^2$, and $J_4$ (\citealt{Veras2007}). The additional secular expression arising from $C_{22}^2$ as derived by \cite{Desaedeleer2006} was also taken into account, while proved to be of a very weak influence on the results. Finally, the solar perturbation $\langle R_{\odot} \rangle$ was modeled by using the analytical expansion of \cite{Brumberg1971} which can be easily averaged over the mean anomalies of the satellite, while retaining the solar orbital evolution in spherical coordinates. The orbit of the Sun was modeled as a precessing orbit, where the three fundamental frequencies ($n_{\odot},\dot{\varpi}_{\odot},\dot{\Omega}_{\odot}$) were taken from the AstDys database (\citealt{Knezevic2003}). 

\begin{figure}[h]
\centering
%\resizebox{12.cm}{!}{\includegraphics [angle=270,width=\textwidth] {carte_seculaire_romulus.ps}}
\resizebox{12.cm}{!}{\includegraphics [angle=0,width=\textwidth] {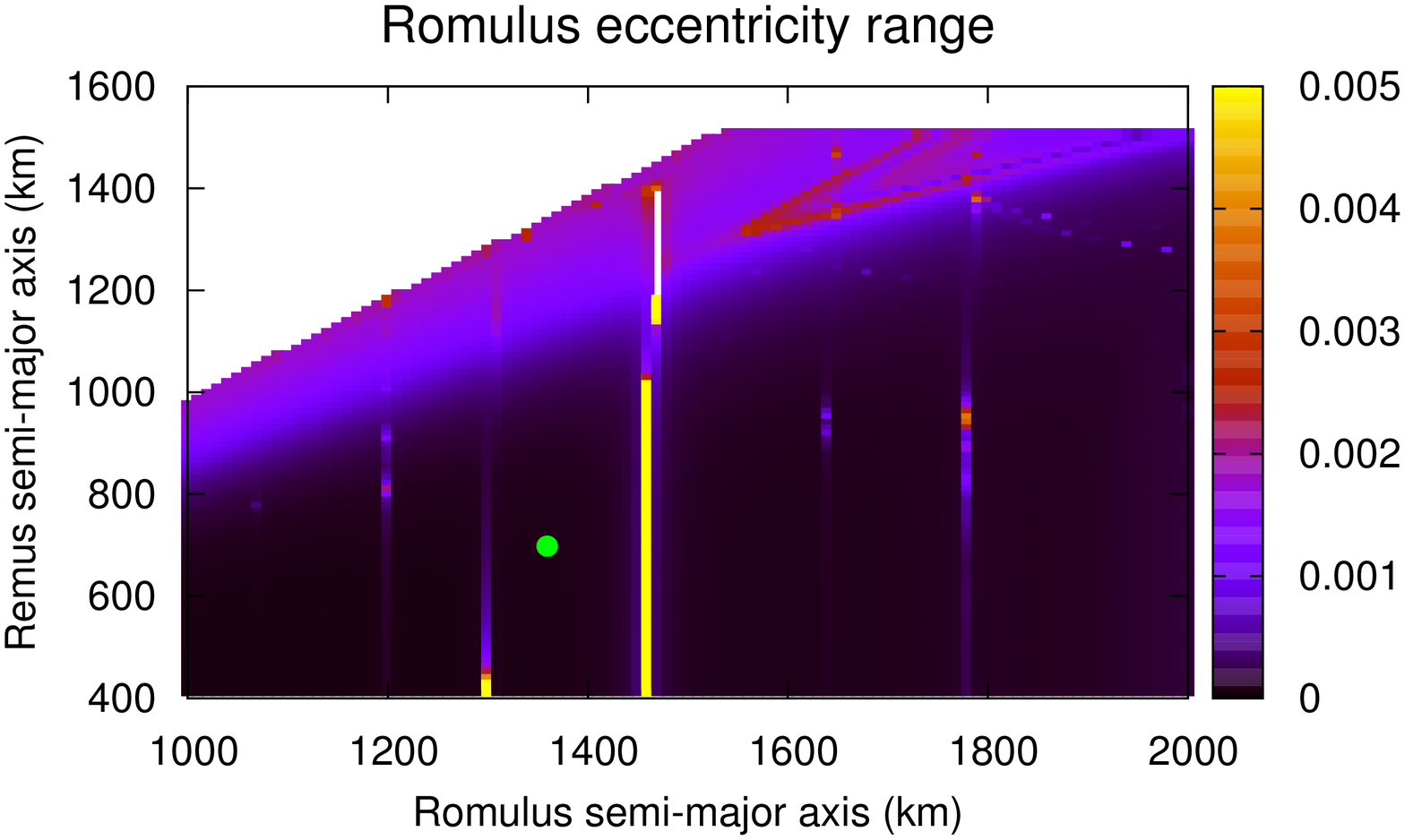}}
\caption{Maximum eccentricity range attained by Romulus over 6600 years, in function of the initial semi-major axis of the satellites. A green point indicates the actual position of the system.}
\label{fig4}
\end{figure}

The use of such averaged disturbing functions implies fixed semi-major axes for the satellites, as well as the suppression of the mean motion resonances from the dynamical system. The same orbits than in Fig.\ref{fig1} (varying the semi-major axis of the satellites) are integrated over 6600 years. The results are shown in Fig.\ref{fig4}.
While the MEGNO could also be used in these integrations (by deriving the Lagrange equations over the orbital elements), we chose here to show the maximum eccentricity range attained by Romulus, which is also an indicator of the chaotic diffusion suffered by the satellite orbits. The white region indicates orbits for which the semi-major axis of Remus exceeds the one of Romulus, and are thus not numerically integrated. The color code range has been limited and chosen in order to magnify the different secular resonances. Vertical lines are visible and represent resonances belonging to the evection family, here between the pericenter frequency of Romulus and the mean motion of the Sun : 
\begin{equation}
k_1 \dot{\varpi} \backsimeq k_2 n_{\odot},
\end{equation} 
with $k_1, k_2$ integers. In particular, the so-called evection resonance ($k_1=k_2=1$) and its implications on the dynamics of satellites have been studied by many authors (\citealt{Touma1998}; \citealt{Breiter2000}; \citealt{Yokoyama2008}; \citealt{Cuk2010}; \citealt{Frouard2010}) and is located in Fig.\ref{fig4} at a Romulus semi-major axis of 1460 km. It is the most distinctive feature in the map. Inside the resonance, the orbit of Romulus becomes highly chaotic, its eccentricity being raised to 0.2. The other resonances from the evection family have a much weaker influence on the orbital evolutions. We stress that the evection resonances considered here are due to the oblateness of the asteroid, which accelerates considerably the precession period of $\varpi$. The evection resonance can also be found much farther in semi-major axis, where the acceleration of the precession period is only due to the Sun.%, and thus close to the stability limit

Despite the fact that the map shows the eccentricity range of Romulus, we can still see some effects due to the evection resonances for Remus. Such resonances are present in the upper right corner of the map; for example one starts at $a_{rom}=2000$ km and $a_{rem}=1250$ km. These resonances should be approximately horizontal in the case where the mass of Remus is much greater than the one of Romulus, but here the perturbation of Romulus sensibly changes the frequency of the pericenter of Remus, with the effect of shaping these resonances as curves. 

Once again, the position of the actual system is shown on the map and lies in a very stable zone.

\section{The effect of tides and BYORP}

We now investigate how tidal and BYORP effects can bring the system through the resonances and what are the timescales involved.

As explained by \cite{Goldreich2009} and \cite{Taylor2010}, the tides raised by the primary on a secondary drive very efficiently the spin of the secondary $\Omega_s$ in synchronization with the mean motion $n$. We thus assume in the following that $\Omega_{Rom}=n_{Rom}$ and $\Omega_{Rem}=n_{Rem}$. 

In a general way, the tidal equations describing the variation of the semi-major axis and eccentricity of a satellite are, for $e \ll 1$ (\citealt{Goldreich1966}; \citealt{Murray2000}) :   
\begin{equation}
\dot{a}^T = sign(\Omega_p - n) \frac{3 k_{2p}}{Q_p} \frac{m_s}{m_p} \bigg( \frac{R_p}{a} \bigg)^5 n a ,
\end{equation} 
\begin{equation}
\dot{e}^T = sign(2 \Omega_p - 3 n) \frac{57}{8} \frac{k_{2p}}{Qp}  \frac{m_s}{m_p} \bigg( \frac{R_p}{a} \bigg)^5 n e ,
\end{equation} 
for the tides raised on the planet by a (prograde) satellite, and  
\begin{equation}
\dot{a}^T = sign(\Omega_s - n) \frac{3 k_{2s}}{Q_s} \frac{m_p}{m_s} \bigg( \frac{R_s}{a} \bigg)^5 n a ,
\end{equation} 
\begin{equation}
\dot{e}^T = - \frac{21}{2} \frac{k_{2s}}{Q_s}  \frac{m_p}{m_s} \bigg( \frac{R_s}{a} \bigg)^5 n e ,
\end{equation} 
for the tides raised on a prograde satellite by a planet. $R$ is the radius, $m$ is the mass and the suffixes $p$ and $s$ represent respectively the primary body and the satellite. Note that the last expression in $\dot{e}^T$ is non-null even if the satellite has its spin synchronized. The equations depend on $k_2$ (which depends on the rigidity of the material) and on the dissipation function $Q$. 

\noindent
Thus, when $\omega_s = n$, the tides raised on the planet by a satellite act to increase both the semi-major axis and the eccentricity, while those raised on the satellite by the planet are negligible for the semi-major axis and decrease the eccentricity, sometimes completely counterbalancing the tides raised on the planet (\citealt{Goldreich1963}). %reference needed
\medskip

\noindent
The tides depend on the parameters $k_2$ and $Q$ for each body, with 
\begin{equation}
k_2 = \frac{3/2}{1 + \tilde{\mu}},
\end{equation}
where $\tilde{\mu}$ is the dimensionless rigidity defined by : 
\begin{equation}
\tilde{\mu} = \frac{19 \mu}{2 g \rho R}.
\end{equation}
$g$ is the gravity at the radius of the body ($g = G m / R^2$), $\rho$ is its density, $\mu$ is the rigidity of the material and $G$ is the gravitational constant. Typical values of the rigidity are $\sim5\times 10^{10} N.m^{-2}$ for a rocky body, or $\sim4\times 10^{9} N.m^{-2}$ for an icy body (\citealt{Murray2000}). \cite{Marchis2008a,Marchis2008b} used $5\times 10^{8} N.m^{-2}$ considering a moderately fractured asteroid for Sylvia, corresponding to $k_{2p} \approx 0.015$. However, \cite{Goldreich2009} showed how the effective rigidity of a rubble-pile can be considerably inferior to a monolith one, and gave the approximation $\tilde{\mu}_{rubble} \sim  (\tilde{\mu}/\epsilon_Y)^{1/2}  $ where $\epsilon_Y$ is the yield strain and is taken to $10^{-2}$. This leads to $k_{2p} \propto R$, although this was infirmed by \cite{Jacobson2011} who found a relation $k_{2p} \propto 1/R$ based on observations of the binary asteroid population.
\medskip

\noindent
The value of the dissipation fonction $Q$ is less known and is often chosen as $\sim 100$ for monoliths (Goldreich \& Sotter 1966). 
\medskip

%\noindent
%We could use as mean radius of the satellites the approximate diameters observed by \cite{Marchis2005}. But we make the assumption that the %satellites originates from Sylvia, assume a common mean density, and then derive mean radius for the satellites by using the density value of %Sylvia and the masses of the satellites, which are much more constrained than their physical radius.
%\medskip

In this work we also take into account the BYORP effect which arises from an asymmetry of the satellite shapes (\citealt{Cuk2005}; \citealt{Mcmahon2010}). We use the following equations from \cite{Jacobson2011} : 
\begin{equation}
\dot{a}^B = \pm \frac{3 H_{\odot} B}{2 \pi} \bigg( \frac{a^{3/2}}{\omega_d \rho R_p^2}  \bigg) \frac{\sqrt{1+m_s/m_p}}{(m_s/m_p)^{1/3}},
\end{equation} 
\begin{equation}
\dot{e}^B = \mp \frac{3 H_{\odot} B}{8 \pi} \bigg( \frac{a^{1/2} e}{\omega_d \rho R_p^2}  \bigg) \frac{\sqrt{1+m_s/m_p}}{(m_s/m_p)^{1/3}},
\end{equation} 
with $\omega_d = \sqrt{4 \pi G \rho /3 }$ and $H_{\odot} = F_{\odot}/(a^2_{\odot} \sqrt{1 - e^2_{\odot}})$ where $F_{\odot}$ is the solar radiation constant. $a_{\odot}$ and $e_{\odot}$ are the semi-major and eccentricity of the Sun. The BYORP effect depends on the parameter $B$ which represents the deviation of the secondary with respect to a symmetric body. $B$ is contained in the interval [0,2] and is commonly taken to $10^{-3}$. As explained in \cite{Jacobson2011}, the sign of $\dot{a}^B$ and $\dot{e}^B$ depend on the shape of the satellites. We assume first $\dot{a}^B>0$ and $\dot{e}^B<0$.

We can derive an approximate tidal age of the satellites by integrating numerically the tidal and BYORP equations backward in time. The satellites are considered separately. The evolution is stopped when the satellites attain the stability limit $a = 400$ km. We show the results for the representative tidal parameters $\{\mu Q,B\} = [10^{11},10^{-3}],[10^{10},10^{-3}],[10^{10},10^{-2}]$ in Table \ref{tab1}.
\begin{table}
\begin{center}
\begin{tabular}{l|lll}
	& $\mu Q$=$10^{10}$, $B$=$10^{-3}$	& $\mu Q$=$10^{10}$, $B$=$10^{-2}$	&$\mu Q$=$10^{11}$, $B$=$10^{-3}$\\
\hline
Remus 	& -47.8 Myr 				&-45.7 Myr				& -453.2 Myr\\
Romulus &-200 Myr  				&-182 Myr  				&	-1.8 Gyr \\
\hline
$a_{rom}$&1299.46 km				&1291.53 km				&1291.57 km
\end{tabular}
\end{center}
\caption{Dynamical ages of the satellites obtained from the tidal and BYORP evolution for different values of the tidal parameters $\mu Q$ and $B$. The satellites are considerated separately. The system is stopped when the satellites attain the semi-major axis limit $a_{crit}=400$ km. The last row indicates the semi-major axis of Romulus when Remus attains $a_{crit}$.}
\label{tab1}
\end{table} 
The tidal age of Romulus is of the same order as the one obtained by Taylor \& Margot (2011) if we consider $\mu Q$ = $10^{11}$. The semi-major axis of Romulus at the time when Remus attains the stability limit is also indicated, which reflects the tidal age of the actual system.

%From Table \ref{tab1}, we can observe that the BYORP effect is more important for distant orbits. 
The BYORP parameter $B$ determines the amplitude of the BYORP effect, but has a reduced effect for the orbits considered here.
\begin{figure}[h]
\centering
\resizebox{12.cm}{!}{\includegraphics [angle=0,width=\textwidth] {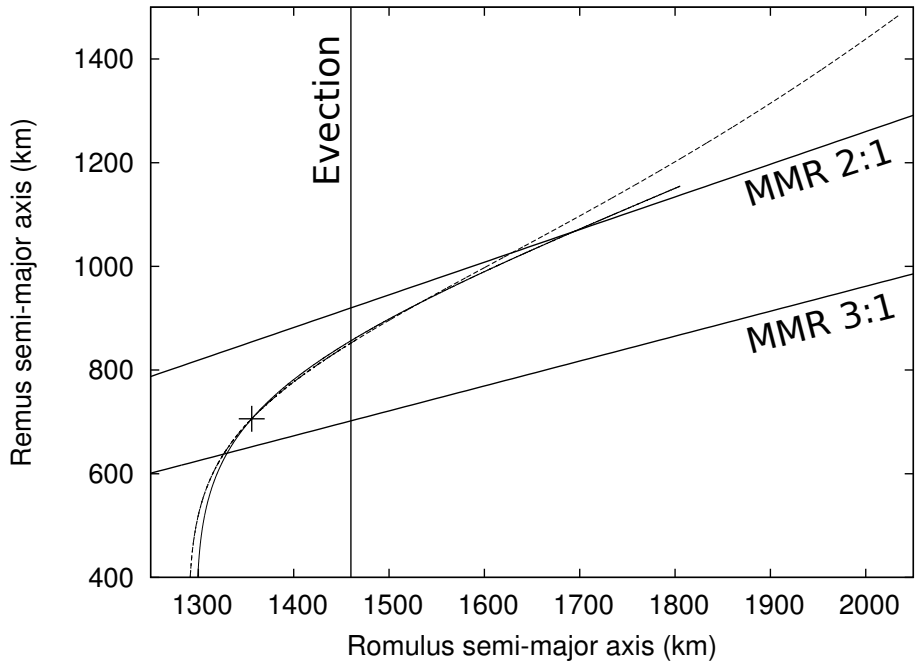}}
\caption{Paths followed by the system due to the effect of tides and BYORP with different values of the parameters $\mu Q$ and $B$. The actual position of the system is indicated by a cross. The solid line corresponds to the parameters $\mu Q$=$10^{10}$, $B$=$10^{-3}$ while the dashed line corresponds to $\mu Q$=$10^{10}$, $B$=$10^{-2}$. The evolution corresponding to the parameters $\mu Q$=$10^{11}$, $B$=$10^{-3}$ is merged with the dashed line. See text for comments.}
\label{fig5}
\end{figure}
We show in Fig.\ref{fig5} the tidal evolution of the system with the three set of parameters. The evolution starts with the semi-major axis of Remus at 400 km up to the actual configuration, and is subsequently followed during 1 Gyr. The major resonances are shown in the figure. The evolution for the parameters $\mu Q$=$10^{11}$, $B$=$10^{-3}$ is merged with the one corresponding with $\mu Q$=$10^{10}$, $B$=$10^{-2}$, and ends shortly after the evection line. The satellites are on converging orbits (the orbit of Remus is expanding faster than the one of Romulus) for the three sets of parameters, and the system crosses the evection resonance before the mean-motion resonance 2:1 in each case.

As observed above, the BYORP effect is stronger for distant orbits. We can see in Fig.\ref{fig5} that a larger value of $B$ allows a faster evolution. It is interesting to note that it also reduces the distance between the two satellites, and thus drives the system towards the mean-motion resonances seen in Fig.\ref{fig1}, and eventually towards highly chaotic zones. A smaller value of $B$ keeps away the system from these zones as the integration time increases. A choice of $B$ = $10^{-2}$ also has the effect of decreasing the eccentricity of Remus after $\sim$ 750 Myr. Its value monotically increases with the two other sets of parameters. 
\medskip

Similarly, using inverse signs for $\dot{a}^B$ and $\dot{e}^B$ noticeably changes the evolutions only when $B$ = $10^{-2}$. For this value, the orbits become divergent after 450 Myr, and the eccentricity of Remus can attain 0.08 after 1 Gyr, compared to 0.03 with our first choice of sign in the BYORP equations. The inverse signs have a very small effect on the dynamical ages of Table \ref{tab1}.
%s $[10^{11},10^{-3}]$ and $ [10^{10},10^{-3}]$.  

Finally, we can observe that the system will also encounter the evection resonance before the MMR 2:1 for the second choice of sign in the BYORP equation and the three set of parameters. We note that with $B$ = $10^{-2}$, the system can even stay between the MMR 2:1 and 3:1 during 5 Gyr.

%*****************************************************
\begin{comment}

\noindent
From the chaotic map (Fig.\ref{fig1}), we know that the orbit of Remus is unstable if $a_{remus} < 400 km$. Expression ? can be used to determine an approximate tidal age for each satellite (Fig.?) by setting $a_i=400km$. By doing so, we assume that no major resonances caused instabilities since the time when $a_{remus} = 400 km$.
\medskip

\noindent
Equations for the evolution of the semi-major axis (??) can be integrated and give (Murray \& Dermott 1999) : 
\begin{equation}
a_o^{13/2} - a_i^{13/2} = \frac{39}{2} \frac{k_{2p}}{Q_p} \bigg(  \frac{G}{m_p} \bigg)^{1/2} R_p^5 m_s  \delta t
\end{equation} 
where $a_o$

We can use this expression to find the time when Romulus will attain the evection resonance at $a=1470 km$ (graph?)
\medskip

\noindent
There is a problem because $e_{Remus}>e_{Romulus}$. If Romulus has been subjected to tides for more time than Remus, its eccentricity should be higher.

\end{comment}
%*****************************************************

\section{Generic instability zones}

As we know now that the orbits of Remus and Romulus are located in a stable dynamical zone, it is interesting to know if it is also the case for the other known triple systems. We also extend this study to satellites of Main-belt asteroids, in the case where the masses of the satellites are significantly smaller than the mass of the asteroid. For greater relative masses of the satellites, one can no longer neglect the influence of their rotation and shape on their orbital evolution (see \citealt{Cuk2010}).

The orbital elements of the asteroids have been taken from the AstDys database. Data concerning the physical parameters of the asteroids and associated satellites, as well as the orbital elements of the satellites, have been taken from various sources; (22) Kalliope (\citealt{Descamps2008}; \citealt{Marchis2008a}), (45) Eugenia (\citealt{Kaasalainen2002}; \citealt{Marchis2010}), (87) Sylvia (\citealt{Marchis2005a}), (93) Minerva (\citealt{Marchis2011}), (107) Camilla (\citealt{Torppa2003}; \citealt{Marchis2005c}), (121) Hermione (\citealt{Marchis2005b}; \citealt{Descamps2009}), (130) Elektra (\citealt{Marchis2008b}; DAMIT), (216) Kleopatra (\citealt{Descamps2011}), (243) Ida (\citealt{Petit1997}), (283) Emma (\citealt{Marchis2008b}; DAMIT), (136108) Haumea (\citealt{Ragozzine2009}; \citealt{Rabinowitz2006}).

\begin{figure}[h]
\centering
%\resizebox{12.cm}{!}{\includegraphics [angle=270,width=\textwidth] {normalized_multiple.ps}}
\resizebox{12.cm}{!}{\includegraphics [angle=0,width=\textwidth] {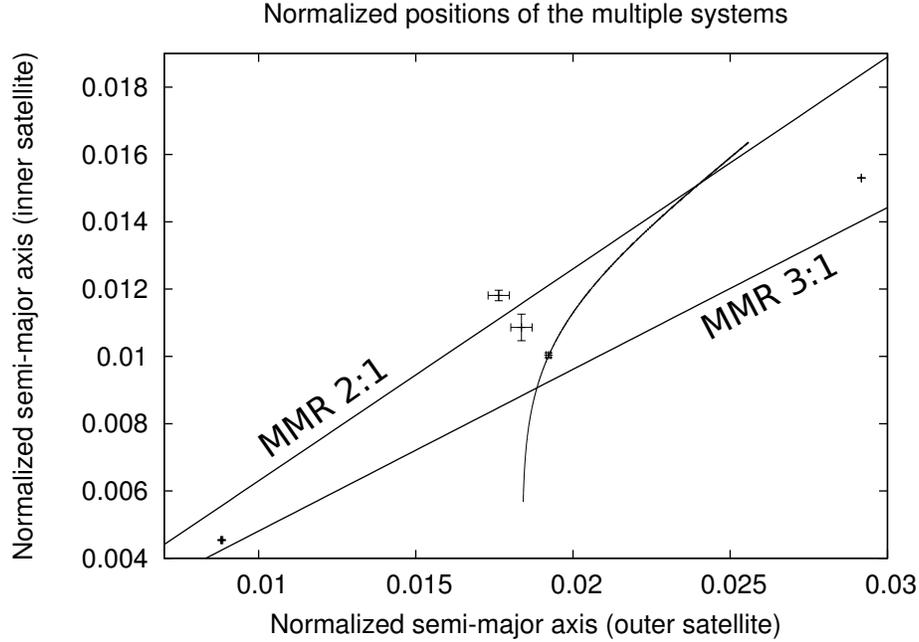}}
%\caption{Normalized semi-major axis of the satellites belonging to multiple systems. }
\caption{Semi-major axis of the multiple systems normalized by the Hill radius of their asteroid. From left to right : (136108) Haumea, (216) Kleopatra, (93) Minerva, (87) Sylvia, (45) Eugenia. One of the possible tidal evolution of (87) Sylvia from Fig.\ref{fig5} is also shown (corresponding to the tidal parameters $\mu Q$ = $10^{10}$, $B$=$10^{-3}$).}
\label{fig6}
\end{figure}

\begin{figure}[h]
\centering
%\resizebox{10.cm}{!}{\includegraphics [angle=270,width=\textwidth] {remus_e_mmr2_1.eps}}
\resizebox{10.cm}{!}{\includegraphics [angle=270,width=\textwidth] {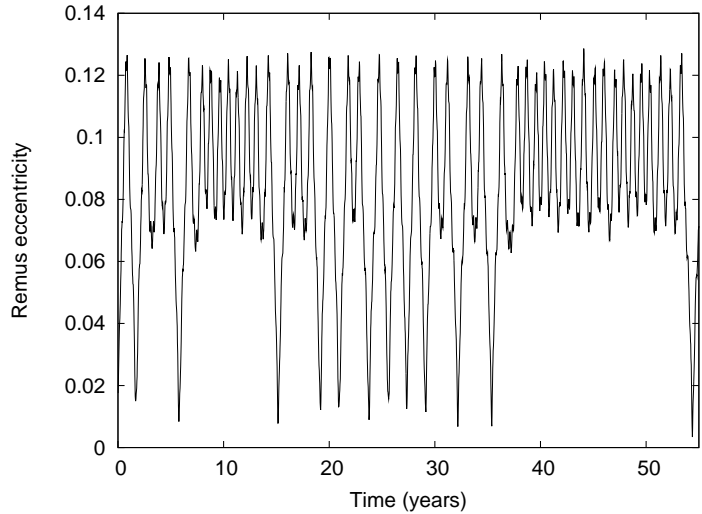}}
\caption{Evolution of the eccentricity of Remus once the system is inside the mean motion resonance 2:1.}
\label{fig8}
\end{figure}

The determination of the location of the triple systems relatively to their mean motion resonances just requires the knowledge of the periods of the satellites. We show in Fig.\ref{fig6} the positions in semi-major axis of the five multiple systems considered here : (45) Eugenia, (87) Sylvia, (93) Minerva, (216) Kleopatra, and (136108) Haumea. The semi-major axes are normalized by the Hill radius of the asteroid, computed with $R_H = a_{ast} (m_{ast}/(3 M_{\odot}))^{1/3}$, where $a_{ast}$ and $m_{ast}$ are the semi-major axis and mass of the asteroid, and $M_{\odot}$ is the solar mass.
The figure shows that all the multiple systems, except (216) Kleopatra, are between the mean motion resonances 2:1 and 3:1. Noting that the tidal and BYORP evolutions move the systems outwards, it could be tempting to postulate that the MMR 2:1 prevents the evolution of the satellites from going beyond its location. In Fig.\ref{fig8}, we show the evolution of the eccentricity of Remus inside the MMR 2:1. The semi-major axes of the satellites are chosen in order to place their orbit inside the resonance according to the tidal evolution shown in Fig.\ref{fig6} ($a_{rem}$=1067 km, $a_{rom}$=1694 km). The other initial elements are chosen equal to their actual values. The eccentricity of Romulus is much less perturbed and attains only 0.01 for the same timespan, and the inclinations of both satellites remain close to their initial values. Using the capture probability estimates of \cite{Dermott1988}, the satellites are assured to be captured in the resonance if their initial eccentricities are less than the critical values $e_{rem}^*=0.094$ and $e_{rom}^*=0.012$.

The actual eccentricities of the satellites could thus indicate whether they suffered a chaotic evolution in the past. %The memory of these events could not be erased since tidal evolutions only increases the eccentricity (preuve). 
(216) Kleopatra is the only system that seems to have crossed the resonance but, unfortunately, we still have no informations about the actual eccentricities of its satellites. 

\cite{Ragozzine2009} studied the orbital and tidal evolution of (136108) Haumea, and argued that the passage of the system through the MMR 3:1 could explain the high value of eccentricity of its inner satellite Namaka ($e=0.249$). However, it can be noted that the known eccentricites of the satellites in the three systems (45) Eugenia, (87) Sylvia and (93) Minerva are quite low; the maximum value is attained by Princess (the innermost satellite of Eugenia) with $e=0.069$. This could indicate than the passage through the MMR 3:1 is quite safe for the satellites. %(figure avec evolution dans la 3:1).
Although the location, dimensions and mass of Haumea make it a completely different object than the Main-Belt asteroids studied here, with possibly different dynamical mecanisms, we can try to find another explanation for these differences in eccentricity. 
\medskip

We show in the following the location of the satellites with respect to the evection resonance. In order to compute the semi-major axis corresponding to the evection $a_{evec}$ for each satellite, we solve $\dot{\varpi}=n_{\odot}$ for $a_{evec}$. The frequency of the pericenter $\dot{\varpi}$ is obtained with the corresponding Lagrange equation : 
\begin{equation}
\dot{\varpi} = \frac{1}{n a^2} \bigg( \frac{\sqrt{1 - e^2}}{e} \frac{\partial \langle R \rangle}{\partial e} + \frac{1 - \cos(I)}{\sqrt{1 - e^2} \sin(I)} \frac{\partial \langle R \rangle}{\partial I} \bigg) 
\label{eq1}
\end{equation} 
where the averaged disturbing function $ \langle R \rangle = \langle R_{obl} \rangle + \langle R_{\odot}\rangle$ contains the solar influence and the oblateness of the asteroid. $\langle R_{obl} \rangle$ is taken from the expansion of \cite{Veras2007} which keeps terms up to $e^4$ and $\sin^4 I$, and $\langle R_{\odot}\rangle$ is the classical Koza\"{\i} approximation. The eccentricity and inclination of the satellite are fixed values, but the solution $a_{evec}$ still depends on the argument of the pericenter $\omega$ of the satellite, although we checked that this dependence is highly negligible. The oblateness coefficients $J_2$ and $J_4$ are determined from the shape of the ellipsoid approximation for each asteroid, using the formulation of \cite{Boyce1997}.

%The mass of the satellites of (216) Kleopatra are determined by considering them as spheres, with the density of Kleopatra and radius given by Marchis et al. (2011). 

An averaged disturbing function describing the perturbation of an additional satellite was used in the case of triple systems. The resulting disturbing function $\langle R_{mut} \rangle$ was taken from the analytical expansion of \cite{Murray2000}. In this case, the value of $\dot{\varpi}$ of each of the satellite was found while keeping fixed the semi-major of the other satellite. In the case of (87) Sylvia and (136108) Haumea, it was found that the frequency of pericenter of the innermost satellite could not be low enough to be commensurable with the mean motion of the Sun. In addition, the resonance can be found for the innermost satellite of (216) Kleopatra only when considering the first shape model, giving two values of $a_{evec}$ : 915 km and 1456 km. However, we did not retain this system in the study, since the lack of information about the eccentricities of its satellites. Similarly, some binary systems have not been retained in this study due to the lack of information about the shape of their asteroid to this date. This is the case of (379) Huenna, (702) Alauda and (762) Pulkova.

For some systems, both the shape of the asteroid from the ellipsoid approximation, and a theorical value of $J_2$ determined from the knowledge of the satellite orbit are known. The computation of $a_{evec}$ gives approximately similar results.  

\begin{figure}[h]
\centering
%\resizebox{12.cm}{!}{\includegraphics [angle=270,width=\textwidth] {normalized_satellites.ps}}
\resizebox{12.cm}{!}{\includegraphics [angle=270,width=\textwidth] {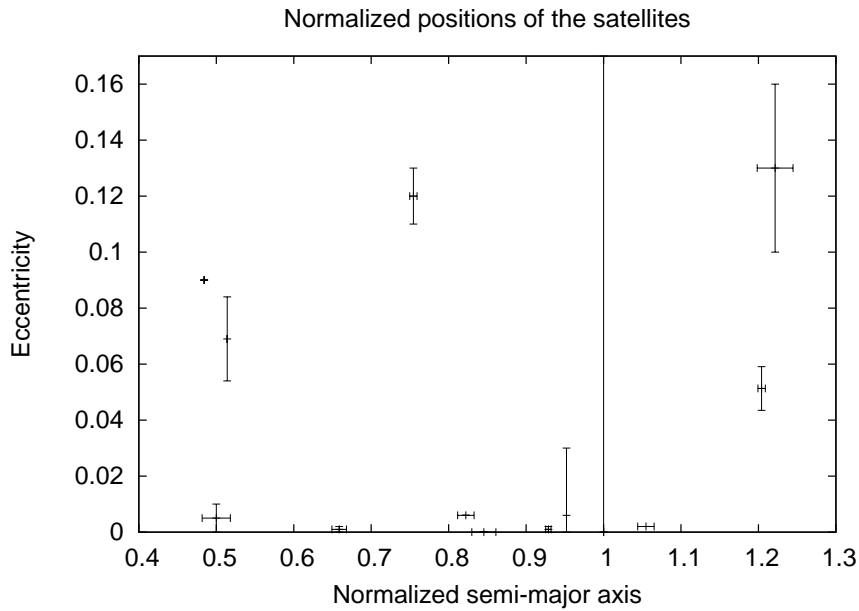}}
\caption{Semi-major axis and eccentricity of the satellites. The semi-major axis is normalized with the position of the evection resonance, represented by the vertical line.}
\label{fig7}
\end{figure}

The distance of the satellites from the evection resonance is shown in Fig.\ref{fig7}. The uncertainties in semi-major axis and eccentricity correspond to the uncertainties on the knowledge of the orbits.  
%Despite the limited number of satellites considered, we can observe three domains. 
The small number of systems available make statistical considerations rather ambiguous but we can try to make a few considerations, based on the assumption that the tidal and BYORP effects make the satellites evolve outwards.

\begin{figure}[h]
\centering
%\resizebox{12.cm}{!}{\includegraphics [angle=270,width=\textwidth] {normalized_satellites.ps}}
\resizebox{12.cm}{!}{\includegraphics [angle=270,width=\textwidth] {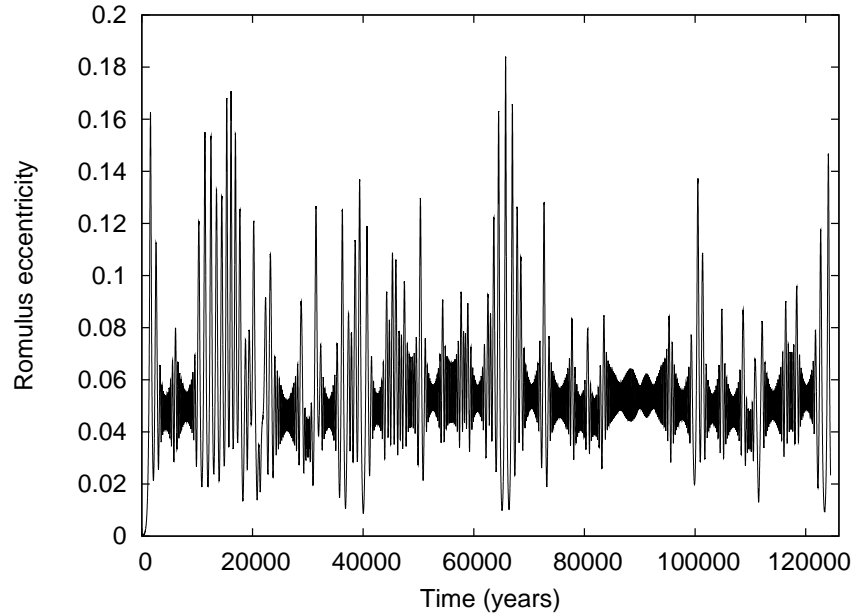}}
\caption{Eccentricity of Romulus inside the evection resonance.}
\label{fig12}
\end{figure}

First we can consider the fact that satellites beyond the evection resonance are rather close to it. For a comparison, the Moon would be located at $\sim 31$ $a_{evec}$ in the figure (using for the Earth $J_2 = 1083\times 10^{-6}$ and $J_4 = -2\times 10^{-6}$). Apart from observational biases, this could maybe point out the evection resonance as a powerful way to eject satellites of asteroids. The effect of the evection resonance is already well-known for dramatically increasing the eccentricity of a satellite (\citealt{Touma1998}; \citealt{Cuk2010}). As an example, we show in Fig.\ref{fig12} the highly chaotic evolution of Romulus inside the resonance. Its initial eccentricity, as well as the initial semi-major axis and eccentricity of Remus, have been given accordingly to the tidal and BYORP evolution of the system, with the parameters $\mu Q$ = $10^{10}$, $B$=$10^{-3}$. 
The larger number of satellites with very-low eccentricity located before the resonance could infer that hypothesis, which will be investigated in a following paper.

\section{Conclusion}

We studied the dynamics and stability of the (87) Sylvia triple system by using numerical integrations of the complete and averaged equations of motion. We used a shape of Sylvia derived from light-curves observations and up to $C_{4,4}, S_{4,4}$ for the complete integrations. The position of the actual system lies in a very stable zone. We showed the possible evolutions of the system trough tidal and BYORP effects, and showed that the system, currently lying between the mean-motion resonances 2:1 and 3:1, will likely evolve through the evection resonance before the MMR 2:1 in the future. The other known triple system considered here, except (216) Kleopatra, also lie between the MMR 2:1 and 3:1. Finally, we show that the evection resonance could limit the outward evolution of the satellites.

\paragraph*{Aknowledgments}
The authors warmly thank A. Lemaitre for her constructive comments and her careful reading of the manuscript. Numerical simulations were made thanks to the local computing resources (Clusters ISCF and URBM-SYSDYN) at the University of Namur (FUNDP, Belgium). This work was supported by FAPESP (process n\degre 2010/52715-5).

\bibliographystyle{model2-names}

\begin{thebibliography}{0}
\expandafter\ifx\csname natexlab\endcsname\relax\def\natexlab#1{#1}\fi
\expandafter\ifx\csname url\endcsname\relax
  \def\url#1{\texttt{#1}}\fi
\expandafter\ifx\csname urlprefix\endcsname\relax\def\urlprefix{URL }\fi
\providecommand{\eprint}[2][]{\url{#2}}
\providecommand{\bibinfo}[2]{#2}
\ifx\xfnm\relax \def\xfnm[#1]{\unskip,\space#1}\fi

\end{thebibliography}


\begin{thebibliography}{00}

\bibitem[{{Bazs{\'o}} et~al.(2010)}]{Bazso2010}
  {{Bazs{\'o}}, {\'A}. and {Dvorak}, R. and {Pilat-Lohinger}, E. and {Eybl}, V. and {Lhotka}, C.},  {A survey of near-mean-motion resonances between Venus and Earth}. {Celestial Mechanics and Dynamical Astronomy}, 107, 63-76, 2010.

\bibitem[{{Belton} et~al.(1996)}]{belton1996}
  {Belton, M.J.S., and 20 colleagues},
  {The discovery and orbit of 1993(243)1 Dactyl}. {Icarus}, 120, 185-199, 1996.

\bibitem[{{Bou{\'e}} \& {Laskar}(2009)}]{Boue2009}
  {{Bou{\'e}}, G. and {Laskar}, J.},
  {Spin axis evolution of two interacting bodies}. {Icarus}, 201, 750-767, 2009.

\bibitem[{{Boyce}(1997)}]{Boyce1997}
  {{Boyce}, W.},  {Comment on a Formula for the Gravitational Harmonic Coefficients of a Triaxial Ellipsoid}. {Celestial Mechanics and Dynamical Astronomy}, 67, 107-110, 1997.

\bibitem[{{Breiter}(2000)}]{Breiter2000}
  {{Breiter}, S.},  {The Prograde C7 Resonance for Earth and Mars Satellite Orbits}. {Celestial Mechanics and Dynamical Astronomy}, 77, 201-214, 2000.

\bibitem[{{Breiter} et al.(2005)}]{Breiter2005}
  {{Breiter}, S. and {Melendo}, B. and {Bartczak}, P. and {Wytrzyszczak}, I.},  {Synchronous motion in the Kinoshita problem. Application to satellites and binary asteroids}. {Astronomy \& Astrophysics}, 437, 753-764, 2005.

\bibitem[{{Brouwer} \& {Clemence}(1961)}]{Brouwer1961}
  {{Brouwer}, D. and {Clemence}, G.~M.},
  {Methods of celestial mechanics}. {New York: Academic Press, 1961}

\bibitem[{{Brumberg} et al.(1971)}]{Brumberg1971}
  {{Brumberg}, V.~A. and {Evdokimova}, L.~S. and {Kochina}, N.~G.},
  {Analytical methods for the orbits of artificial satellites of the moon}. {Celestial Mechanics and Dynamical Astronomy}, 3, 197-221, 1971

\bibitem[{{Chauvineau} et~al.(1993)}]{Chauvineau1993}
  {{Chauvineau}, B. and {Farinella}, P. and {Mignard}, F.},
  {Planar orbits about a triaxial body - Application to asteroidal satellites}. {Icarus}, 105, 370, 1993.

\bibitem[{{Comp\`ere} et~al.(2012a)}]{Compere2012a}
  {{Comp\`ere}, A., {Lema{\^i}tre}, A., \& {Delsate}, N.},
  {Detection by MEGNO of the gravitational resonances between a rotating ellipsoid and a point mass satellite}, Celestial Mechanics and Dynamical Astronomy, 112, 75-98, 2012a.

\bibitem[{{Comp\`ere} et~al.(2012b)}]{Compere2012b}
 {{Comp\`ere}, A. and {Frouard}, J and {Carry}, B}, Approximation of the gravitational potential of a non-spherical object: application to binary and triple asteroid systems, Proceedings of the workshop Orbital couples: "Pas de deux", 2012b.

\bibitem[{{Cincotta} \& {Sim\'{o}}(2000)}]{Cincotta2000}
  {{Cincotta}, P.~M. and {Sim\'{o}}, C.}, {Simple tools to study global dynamics in non-axisymmetric galactic potentials - I}. {{Astronomy \& Astrophysics}, Supplement}, 147, 205-228, 2000.

\bibitem[{{Cincotta} et~al.(2003)}]{Cincotta2003}
  {{Cincotta}, P.~M., {Giordano}, C.~M., \& {Sim\'{o}}, C.}, 
  {Phase space structure of multi-dimensional systems by means of the mean exponential growth factor of nearby orbits}. {Physica D}, 182, 151-178, 2003.

\bibitem[{{{\'C}uk} \& {Burns}(2005)}]{Cuk2005}
  {{{\'C}uk}, M. and {Burns}, J.~A.}, 
  {Effects of thermal radiation on the dynamics of binary NEAs}. {Icarus}, 176, 418-431, 2005.

\bibitem[{{{\'C}uk} \& {Nesvorn{\'y}}(2010)}]{Cuk2010}
  {{{\'C}uk}, M. and {Nesvorn{\'y}}, D.}, 
  {Orbital evolution of small binary asteroids}. {Icarus}, 207, 732-743, 2010.

\bibitem[{{Delsate} et~al.(2010)}]{Delsate2010}
  {{Delsate}, N., {Robutel}, P., {Lema{\^i}tre}, A., \& {Carletti}, T.},
  {Frozen orbits at high eccentricity and inclination: application to Mercury orbiter}. {Celestial Mechanics and Dynamical Astronomy}, 108, 275-300, 2010.

\bibitem[{{Delsate}(2011)}]{Delsate2011}
  {{Delsate}, N.},
  {Analytical and numerical study of the ground-track resonances of {Dawn} orbiting {Vesta}}. {Planetary and Space Science}, 59, 1372-1383, 2011.

\bibitem[{{Delsate} \& {Comp\`ere}(2011)}]{NIMASTEP}
  {{Delsate}, N. and {Comp\`ere}, A.},
  {{NIMASTEP}: a software to modelize, study and analyze the dynamics of various small objects orbiting specific bodies}, submitted in Astronomy \& Astrophysics, 2011.

\bibitem[{{Dermott} et al.(1988)}]{Dermott1988}
  {{Dermott}, S.~F. and {Malhotra}, R. and {Murray}, C.~D.}, {Dynamics of the Uranian and Saturnian satellite systems - A chaotic route to melting Miranda?}. {Icarus}, 76, 295-334, 1988.

\bibitem[{{De Saedeleer} \& {Henrard}(2006)}]{Desaedeleer2006}
  {{De Saedeleer}, B. and {Henrard}, J.},
  {The combined effect of J$_{2}$ and C$_{22}$ on the critical inclination of a lunar orbiter}, {Advances in Space Research}, 37, 80-87, 2006.

\bibitem[{{Descamps} et al.(2008)}]{Descamps2008}
  {{Descamps}, P. and 18 colleagues},
  {New determination of the size and bulk density of the binary Asteroid 22 Kalliope from observations of mutual eclipses}, {Icarus}, 196, 578-600, 2008.

\bibitem[{{Descamps} et~al.(2009)}]{Descamps2009}
{{Descamps}, P. and 31 colleagues}, {New insights on the binary Asteroid 121 Hermione}, Icarus, 203, 88-101, 2009.

\bibitem[{{Descamps} et~al.(2011)}]{Descamps2011}
{{Descamps}, P. and 18 colleagues}, {Triplicity and physical characteristics of Asteroid (216) Kleopatra}, Icarus, 211, 1022-1033, 2011.

\bibitem[{{Drummond} \& Christou(2008)}]{Drummond2008}
{{Drummond}, J. and {Christou}, J.}, {Triaxial ellipsoid dimensions and rotational poles of seven asteroids from Lick Observatory adaptive optics images, and of Cere}, Icarus, 197, 480-49, 2008.

\bibitem[{{Durech} et~al.(2010)}]{DAMIT}
  {{Durech}, J. and {Sidorin}, V. and {Kaasalainen}, M.}, {DAMIT: a database of asteroid models}, {Astronomy \& Astrophysics}, 513, A46, 2010.

\bibitem[{{Fahnestock} \& Scheeres(2008)}]{Fahnestock2008}
  {{Fahnestock}, E.~G. and {Scheeres}, D.~J.}, {Simulation and analysis of the dynamics of binary near-Earth Asteroid (66391) 1999 KW4}, {Icarus}, 194, 410-435, 2008.

\bibitem[{{Frouard} et~al.(2010)}]{Frouard2010}
{{Frouard}, J. and {Fouchard}, M. and {Vienne}, A.}, {About the dynamics of the evection resonance}, {Astronomy \& Astrophysics}, 515, A54, 2010 

\bibitem[{{Frouard} et~al.(2011)}]{Frouard2011}
{{Frouard}, J. and {Vienne}, A. and {Fouchard}, M.}, {The long-term dynamics of the Jovian irregular satellites}, {Astronomy \& Astrophysics}, 532, A44, 2011 

\bibitem[{{Goldreich}(1963)}]{Goldreich1963}
{{Goldreich}, P.}, {On the eccentricity of satellite orbits in the solar system}, {Monthly Notices of the Royal Astronomical Society}, 126, 257, 1963

\bibitem[{{Goldreich} \& {Sari}(2009)}]{Goldreich2009}
{{Goldreich}, P. and {Sari}, R.}, {Tidal Evolution of Rubble Piles}, {Astronomy \& Astrophysics}, 691, 54-60, 2009

\bibitem[{{Goldreich} \& {Soter}(1966)}]{Goldreich1966}
{{Goldreich}, P. and {Soter}, S.}, {Q in the Solar System}, {Icarus}, 5, 375-389, 1966

\bibitem[{{Go\'{z}dziewski} et~al.(2001)}]{Gozdziewski2001}
  {{Go\'{z}dziewski}, K., {Bois}, E., {Maciejewski}, A.~J., \& {Kiseleva-Eggleton}, L.}, {Global dynamics of planetary systems with the \texttt{MEGNO} criterion}. {Astronomy \& Astrophysics}, 378, 569-586, 2001.

\bibitem[{{Go\'{z}dziewski} et~al.(2008a)}]{Gozdziewski2008a}
  {{Go\'{z}dziewski}, K., {Breiter}, S., \& {Borczyk} W.}, {The long-term stability of extrasolar system HD37154. numerical study of resonance effects}. {Monthly Notices of the Royal Astronomical Society}, 383, 989-999, 2008a.

\bibitem[{{Go\'{z}dziewski} et~al.(2008b)}]{Gozdziewski2008b}
  {{Go{\'z}dziewski}, K. and {Migaszewski}, C. and {Konacki}, M.}, {A dynamical analysis of the 14 Herculis planetary system}. {Monthly Notices of the Royal Astronomical Society}, 385, 957-966, 2008b.

\bibitem[{{Guzzo}(2005)}]{Guzzo2005}
{{Guzzo}, M.},{The web of three-planet resonances in the outer Solar System}. Icarus, 174, 273-284, 2005

\bibitem[{{Guzzo}(2006)}]{Guzzo2006}
{{Guzzo}, M.},{The web of three-planet resonances in the outer Solar System. II. A source of orbital instability for Uranus and Neptune}. Icarus, 181, 475-485, 2006

\bibitem[{{Jacobson} \& {Scheeres}(2011)}]{Jacobson2011}
{{Jacobson}, S.~A. and {Scheeres}, D.~J.},{Long-term Stable Equilibria for Synchronous Binary Asteroids}. {The Astrophysical Journal Letters}, 736, L19, 2011

\bibitem[{{Kaasalainen} et~al.(2002)}]{Kaasalainen2002}
{{Kaasalainen}, M. and {Torppa}, J. and {Piironen}, J.}, {Models of Twenty Asteroids from Photometric Data}, Icarus, 159, 369-395, 2002.

\bibitem[{{Kaula}(1966)}]{Kaula1966}
  {Kaula}, W.M., 1966. Theory of satellite geodesy.
  Blaisdell Publishing Company, Waltham Massachusetts, Toronto, London. 

\bibitem[{{Kne{\v z}evi{\'c}} \& {Milani}(2003)}]{Knezevic2003}
{{Kne{\v z}evi{\'c}}, Z. and {Milani}, A.}, {Proper element catalogs and asteroid families}, {Astronomy \& Astrophysics}, 403, 1165-1173, 2003.

\bibitem[{{Kozai}(1959)}]{Kozai1959}
{{Kozai}, Y.}, {The motion of a close earth satellite}, {The Astronomical Journal}, 64, 367-377, 1959.

\bibitem[{{Kozai}(1962)}]{Kozai1962}
{{Kozai}, Y.}, {Second-order solution of artifical satellite theory without air drag}, {The Astronomical Journal}, 67, 446-461, 1962.

\bibitem[{{Laskar}(1990)}]{Laskar1990}
{{Laskar}, J.}, {The chaotic motion of the solar system - A numerical estimate of the size of the chaotic zones}, {Icarus}, 88, 266-291, 1990.

\bibitem[{{Laskar}(2005)}]{Laskar2005}
{{Laskar}, J.}, {Frequency map analysis and quasi periodic decompositions.}, In: Benest, D., Froeschle, C., Lega, E. (Eds.), Hamiltonian Systems and Fourier Analysis. Taylor and Francis. 2005.
      
\bibitem[{{Lema{\^i}tre} et~al.(2009)}]{Lemaitre2009}
  {{Lema{\^i}tre}, A., {Delsate}, N., \& {Valk}, S.},
  {A web of secondary resonances for large A/m geostationary debris}. {Celestial Mechanics and Dynamical Astronomy}, 104, 383-402, 2009.

\bibitem[{{Maciejewski}(1995)}]{Maciejewski1995}
  {{Maciejewski}, A.~J.},
  {Reduction, Relative Equilibria and Potential in the Two Rigid Bodies Problem}. {Celestial Mechanics and Dynamical Astronomy}, 63, 1-28, 1995.

\bibitem[{{Marchis} et~al.(2005a)}]{Marchis2005a}
{{Marchis}, F. and {Descamps}, P. and {Hestroffer}, D. and {Berthier}, J.}, {Discovery of the triple asteroidal system 87 Sylvia}, Nature, 436, 822-824, 2005a.

\bibitem[{{Marchis} et~al.(2005b)}]{Marchis2005b}
{{Marchis}, F. and {Hestroffer}, D. and {Descamps}, P. and {Berthier}, J. and 
	{Laver}, C. and {de Pater}, I.}, {Mass and density of Asteroid 121 Hermione from an analysis of its companion orbit}, Icarus, 178, 450-464, 2005b.

\bibitem[{{Marchis} et~al.(2005c)}]{Marchis2005c}
{{Marchis}, F. and {Berthier}, J. and {Descamps}, P. and {Hestroffer}, D. and {Vachier} F.}, {103 Camilla and S/2001 (103) 1}, Univ. of California-Berkeley Astronomy Dept., on line: \url{http://astro.berkeley.edu/~fmarchis/Science/Asteroids/Camilla.html}, 2005c.

\bibitem[{{Marchis} et~al.(2006)}]{Marchis2006}
{{Marchis}, F. and {Kaasalainen}, M. and {Hom}, E.~F.~Y. and {Berthier}, J. and {Enriquez}, J. and {Hestroffer}, D. and {Le Mignant}, D. and 
{de Pater}, I.}, Shape, size and multiplicity of main-belt asteroids. I. Keck Adaptive Optics survey, Icarus, 185, 39-63, 2006.

\bibitem[{{Marchis} et~al.(2008a)}]{Marchis2008a}
{{Marchis}, F. and {Descamps}, P. and {Baek}, M. and {Harris}, A.~W. and 
	{Kaasalainen}, M. and {Berthier}, J. and {Hestroffer}, D. and 
	{Vachier}, F.}, {Main belt binary asteroidal systems with circular mutual orbits}, Icarus, 196, 97-118, 2008a.

\bibitem[{{Marchis} et~al.(2008b)}]{Marchis2008b}
{{Marchis}, F. and {Descamps}, P. and {Berthier}, J. and {Hestroffer}, D. and 
	{Vachier}, F. and {Baek}, M. and {Harris}, A.~W. and {Nesvorn{\'y}}, D.}, {Main belt binary asteroidal systems with eccentric mutual orbits}, Icarus, 195, 295-316, 2008b.

\bibitem[{{Marchis} et~al.(2010)}]{Marchis2010}
{{Marchis}, F. and {Lainey}, V. and {Descamps}, P. and {Berthier}, J. and 
	{van Dam}, M. and {de Pater}, I. and {Macomber}, B. and {Baek}, M. and 
	{Le Mignant}, D. and {Hammel}, H.~B. and {Showalter}, M. and 
	{Vachier}, F.}, {A dynamical solution of the triple asteroid system (45) Eugenia}, Icarus, 210, 635-643, 2010.

\bibitem[{{Marchis} et~al.(2011)}]{Marchis2011}
{{Marchis}, F. and {Descamps}, P. and {Dalba}, P. and 
	{Enriquez}, J.E. and {Durech}, J. and {Emery}, J.P. and {Berthier}, J. and 
	{Vachier}, F. and {Melbourne}, J. and {Stockton}, A.N. and 
	{Fassnacht}, C.D. and {Dupuy} T.J.}, {A detailed picture of the (93) Minerva triple system}, EPSC-DPS 2011 653-3.

\bibitem[{{Margot} et~al.(2002)}]{Margot2002}
{{Margot}, J.~L. and {Nolan}, M.~C. and {Benner}, L.~A.~M. and 
	{Ostro}, S.~J. and {Jurgens}, R.~F. and {Giorgini}, J.~D. and 
	{Slade}, M.~A. and {Campbell}, D.~B.}, {Binary Asteroids in the Near-Earth Object Population}. {Science}, 296, 1445-1448, 2002.

\bibitem[{{Mathis} \& {Le Poncin-Lafitte}(2009)}]{Mathis2009}
{{Mathis}, S. and {Le Poncin-Lafitte}, C.}, {Tidal dynamics of extended bodies in planetary systems and multiple stars}. {Astronomy \& Astrophysics}, 497, 889-910, 2009.

\bibitem[{{McMahon} \& {Scheeres}(2010)}]{Mcmahon2010}
{{McMahon}, J. and {Scheeres}, D.}, {Detailed prediction for the BYORP effect on binary near-Earth Asteroid (66391) 1999 KW4 and implications for the binary population}. {Icarus}, 209, 494-509, 2010.

\bibitem[{{Milani} \& {Kne{\v z}evi{\'c}}(1998)}]{Milani1998}
{{Milani}, A. and {Kne{\v z}evi{\'c} }, Z.}, {Asteroid Mean Elements: Higher Order and Iterative Theories}. {Celestial Mechanics and Dynamical Astronomy}, 71, 55-78, 1998.

\bibitem[{{Murray} \& {Dermott}(2000)}]{Murray2000}
{{Murray}, C.~D. and {Dermott}, S.~F.}, {Solar system dynamics}. Cambridge University Press, 2000

\bibitem[{{Mysen} et al.(2006)}]{Mysen2006}
{{Mysen}, E. and {Olsen}, {\O}. and {Aksnes}, K.}, {Chaotic gravitational zones around a regularly shaped complex rotating body}. {Planetary And Space Science}, 54, 750-760, 2006.

\bibitem[{{Mysen} \& {Aksnes}(2007)}]{Mysen2007}
{{Mysen}, E. and {Aksnes}, K.}, {On the dynamical stability of the Rosetta orbiter. II.}. {Astronomy \& Astrophysics}, 470, 1193-1199, 2007.

\bibitem[{{Petit} et al.(1997)}]{Petit1997}
{{Petit}, J.-M. and {Durda}, D.~D. and {Greenberg}, R. and {Hurford}, T.~A. and 
	{Geissler}, P.~E.}, {The Long-Term Dynamics of Dactyl's Orbit}. {Icarus}, 130, 177-197, 1997.

\bibitem[{{Pravec} et~al.(2006)}]{Pravec2006}
{{Pravec}, P. and 56 colleagues}, {Photometric survey of binary near-Earth asteroids}. {Icarus}, 181, 63-93, 2006.

\bibitem[{{Pravec} \& Harris(2007)}]{Pravec2007}
{{Pravec}, P. and {Harris}, A.~W.}, {Binary asteroid population. 1. Angular momentum content}. {Icarus}, 190, 250-259, 2007.

\bibitem[{{Pravec} et al.(2011)}]{Pravec2011}
{{Pravec}, P., and 41 colleagues}, {Binary Asteroid Population. 2. Anisotropic distribution of orbit poles}. {Icarus}, submitted.

\bibitem[{{Rabinowitz} et al.(2006)}]{Rabinowitz2006}
{{Rabinowitz}, D.~L. and {Barkume}, K. and {Brown}, M.~E. and 
	{Roe}, H. and {Schwartz}, M. and {Tourtellotte}, S. and {Trujillo}, C.}, {Photometric Observations Constraining the Size, Shape, and Albedo of 2003 EL61, a Rapidly Rotating, Pluto-sized Object in the Kuiper Belt}. {The Astrophysical Journal}, 639, 1238-1251, 2006

\bibitem[{{Ragozzine} \& {Brown}(2009)}]{Ragozzine2009}
{{Ragozzine}, D. and {Brown}, M.~E.}, {Orbits and Masses of the Satellites of the Dwarf Planet Haumea (2003 EL61)}. {The Astronomical Journal}, 137, 4766-4776, 2009

\bibitem[{{Scheeres}(1994)}]{Scheeres1994}
{{Scheeres}, D.~J.}, {Dynamics about uniformly rotating triaxial ellipsoids: Applications to asteroids}. {Icarus}, 110, 225-238, 1994.

\bibitem[{{Scheeres} et al.(1996)}]{Scheeres1996}
{{Scheeres}, D.~J. and {Ostro}, S.~J. and {Hudson}, R.~S. and 
	{Werner}, R.~A.}, {Orbits Close to Asteroid 4769 Castalia}. {Icarus}, 121,67-87, 1996.

\bibitem[{{Scheeres}(2002)}]{Scheeres2002}
{{Scheeres}, D.~J.}, {Stability of Binary Asteroids}. {Icarus}, 159, 271-283, 2002.

\bibitem[{{Steinberg} \& {Sari}(2011)}]{Steinberg2011}
{{Steinberg}, E. and {Sari}, R.}, {Binary YORP Effect and Evolution of Binary Asteroids}. {The Astronomical Journal}, 141, 55, 2011.

\bibitem[{{Taylor} \& {Margot}(2010)}]{Taylor2010}
{{Taylor}, P.~A. and {Margot}, J.-L.}, {Tidal evolution of close binary asteroid systems}. {Celestial Mechanics and Dynamical Astronomy}, 108, 315-338, 2010.

\bibitem[{{Taylor} \& {Margot}(2011)}]{Taylor2011}
{{Taylor}, P.~A. and {Margot}, J.-L.}, {Binary asteroid systems: Tidal end states and estimates of material properties}. {Icarus}, 212, 661-676, 2011.

\bibitem[{{Torppa} et al.(2003)}]{Torppa2003}
{{Torppa}, J. and {Kaasalainen}, M. and {Michalowski}, T. and 
	{Kwiatkowski}, T. and {Kryszczy{\'n}ska}, A. and {Denchev}, P. and 
	{Kowalski}, R.}, {Shapes and rotational properties of thirty asteroids from photometric data}. {Icarus}, 164, 346-383, 2003.

\bibitem[{{Touma} \& {Wisdom}(1998)}]{Touma1998}
{{Touma}, J. and {Wisdom}, J.}, {Resonances in the Early Evolution of the Earth-Moon System}. {The Astronomical Journal}, 115, 1653-1663, 1998.

\bibitem[{{Valk} et~al.(2009)}]{Valk2009}
  {{Valk}, S., {Delsate}, N., {Lema{\^i}tre}, A., \& {Carletti}, T.},
  {Global dynamics of high area-to-mass ratios GEO space debris by means of the MEGNO indicator}. {Advances in Space Research}, 43, 1509-1526, 2009a.

\bibitem[{{Veras}(2007)}]{Veras2007}
  {{Veras}, D.},  {A resonant-term-based model including a nascent disk, precession, and oblateness: application to GJ 876}. {Celestial Mechanics and Dynamical Astronomy}, 99, 197-243, 2007

\bibitem[{{Winter} et~al.(2009)}]{Winter2009}
  {{Winter}, O.~C. and {Boldrin}, L.~A.~G. and {Vieira Neto}, E. and 
	{Vieira Martins}, R. and {Giuliatti Winter}, S.~M. and {Gomes}, R.~S. and 
	{Marchis}, F. and {Descamps}, P.},
  {On the stability of the satellites of asteroid 87 Sylvia}. {Monthly Notices of the Royal Astronomical Society}, 395, 218-227, 2009.

\bibitem[{{Yokoyama} et~al.(2008)}]{Yokoyama2008}
  {{Yokoyama}, T., {Vieira Neto}, E., {Winter}, O.C., {Sanchez}, D.M., {de Oliveira Brasil}, P.I.},
  {On the evection resonance and its connection to the stability of outer satellites}. {Mathematical Problems in Engineering}, doi:10.1155/2008/251978. 2008





%% \bibitem must have one of the following forms:
%%   \bibitem[Jones et al.(1990)]{key}...
%%   \bibitem[Jones et al.(1990)Jones, Baker, and Williams]{key}...
%%   \bibitem[Jones et al., 1990]{key}...
%%   \bibitem[\protect\citeauthoryear{Jones, Baker, and Williams}{Jones
%%       et al.}{1990}]{key}...
%%   \bibitem[\protect\citeauthoryear{Jones et al.}{1990}]{key}...
%%   \bibitem[\protect\astroncite{Jones et al.}{1990}]{key}...
%%   \bibitem[\protect\citename{Jones et al., }1990]{key}...
%%   \harvarditem[Jones et al.]{Jones, Baker, and Williams}{1990}{key}...
%%


\end{thebibliography}

%% not want to use model2-names.bst.


\end{document}